\documentclass[lettersize,journal]{IEEEtran}
\usepackage{amsmath,amsfonts}
\usepackage{algorithmic}
\usepackage{algorithm}
\usepackage{array}
\usepackage{textcomp}
\usepackage{stfloats}
\usepackage{url}
\usepackage{verbatim}
\usepackage{graphicx}
\usepackage{cite}
\usepackage{lipsum}
\usepackage{xspace}
\usepackage{caption}
\usepackage{subcaption}
\usepackage{attachfile2}
\usepackage{multicol}
\usepackage{multirow}
\usepackage{booktabs}
\usepackage{amssymb}
\usepackage{pifont}
\usepackage{hyperref}
\usepackage{xcolor}
\usepackage{marvosym}
\usepackage{arydshln}
\usepackage{soul}
\usepackage{musicography}

\newcommand{\xmark}{\ding{55}}

\def\framework{XMusic\xspace}
\def\projector{XProjector\xspace}
\def\composer{XComposer\xspace}
\def\dataset{XMIDI\xspace}
\newcommand{\MYhref}[3][blue]{\href{#2}{\color{#1}{#3}}}%

\begin{document}

\title{\framework: Towards a Generalized and Controllable Symbolic Music Generation Framework}

\author{Sida Tian$^{\,\musSixteenth}$, Can Zhang$^{\, \musSixteenth}$, Wei Yuan$^{\,\musSixteenth}$, Wei Tan, Wenjie Zhu

\thanks{All the authors are with Tencent. $\,\musSixteenth$: Equal contribution.}}

\maketitle

\begin{abstract}
In recent years, remarkable advancements in artificial intelligence-generated content (AIGC) have been achieved in the fields of image synthesis and text generation, generating content comparable to that produced by humans. However, the quality of AI-generated music has not yet reached this standard, primarily due to the challenge of effectively controlling musical emotions and ensuring high-quality outputs.
This paper presents a generalized symbolic music generation framework, \framework, which supports flexible prompts (\textit{i.e.}, images, videos, texts, tags, and humming) to generate emotionally controllable and high-quality symbolic music. 
\framework consists of two core components, \projector and \composer. 
\projector parses the prompts of various modalities into symbolic music elements (\textit{i.e.}, emotions, genres, rhythms and notes) within the projection space to generate matching music. 
\composer contains a Generator and a Selector. The Generator generates emotionally controllable and melodious music based on our innovative symbolic music representation, whereas the Selector identifies high-quality symbolic music by constructing a multi-task learning scheme involving quality assessment, emotion recognition, and genre recognition tasks. 
In addition, we build \dataset, a large-scale symbolic music dataset that contains 108,023 MIDI files annotated with precise emotion and genre labels. 
Objective and subjective evaluations show that \framework significantly outperforms the current state-of-the-art methods with impressive music quality.
Our \framework has been awarded as one of the nine \textit{Highlights of Collectibles at WAIC 2023}.
The project homepage of XMusic is: \MYhref{https://xmusic-project.github.io}{https://xmusic-project.github.io}.
\end{abstract}


\section{Introduction}
\IEEEPARstart{A}{rtificial} intelligence (AI) techniques have significantly advanced the field of AI Generated Content (AIGC), making it a prominent research area in recent years. AIGC fosters creativity, exploration, and innovation across diverse artistic domains. As an art form centered on sound, music is a significant component of AIGC. Automatic music generation has numerous potential applications, including adaptive soundtracks, video background music generation, music transcription, and royalty-free music creation, \textit{etc}. Although recent studies (such as AudioLM~\cite{borsos2023audiolm}, MusicLM~\cite{agostinelli2023musiclm}, Riffusion~\cite{Forsgren_Martiros_2022}, MusicGen~\cite{copet2023simple} and Noise2Music~\cite{huang2023noise2music}, \textit{etc}) have achieved success in terms of generating music within the \textit{audio domain}, editing such generated music in its audio format remains challenging and unintuitive. In contrast, symbolic music, typically represented in MIDI format, offers greater flexibility, enabling users to modify specific musical elements explicitly. Thus, this paper focuses on music generation within the \textit{symbolic domain}.

\begin{figure}[t]
\begin{center}
\includegraphics[width=1\linewidth]{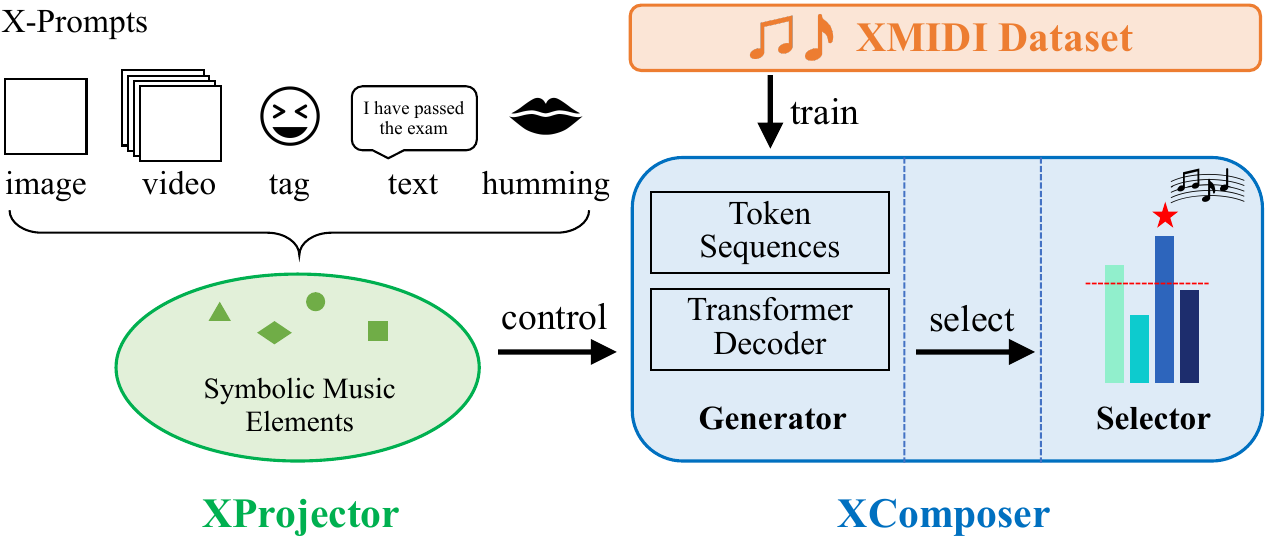}
\end{center}
 \caption{The architectural overview of our \framework framework. It contains two essential components: \projector and \composer. \projector parses various input prompts into specific symbolic music elements. These elements then serve as control signals, guiding the music generation process within the Generator of \composer. Additionally, \composer includes a Selector that evaluates and identifies high-quality generated music. The Generator is trained on our large-scale dataset, \dataset, which includes precise emotion and genre labels.}
\label{fig:intro}
\end{figure}

Symbolic music generation methods primarily aim to model the temporal dependencies within music, predicting subsequent musical events based on prior ones. The Transformer is a natural fit for this sequence-to-sequence task while handling long-range dependencies. Recent advancements have demonstrated the remarkable potential of Transformer models~\cite{vaswani2017attention,dai2019transformer} in symbolic music generation. The Music Transformer by Huang \textit{et al.}~\cite{huangmusic} shows the first successful application of the self-attention mechanism for generating long symbolic music. The Pop Music Transformer~\cite{huang2020pop} incorporates the Transformer-XL~\cite{dai2019transformer} architecture to generate symbolic pop music with an enhanced rhythmic structure. Another influential contribution is the Compound Word Transformer~\cite{hsiao2021compound}, which explores novel and efficient tokenization techniques for symbolic music training.

% \IEEEpubidadjcol

Conditional symbolic music generation has attracted significant attention due to its ability to leverage user-supplied information as a ``prompt'' for producing unique musical compositions. Existing conditional methods explore various types of prompts, including attribute tags (\textit{e.g.}, emotions~\cite{EMOPIA,bao2022generating}, genres~\cite{payne2019musenet}, and instruments~\cite{sarmento2023gtr}), sequential data (\textit{e.g.}, lead sheets~\cite{EmoMusicTV}, motifs~\cite{Zou2021MelonsGM}, and melodies~\cite{huangmusic}), and multimedia data (\textit{e.g.}, performance footage~\cite{gan2020foley,su2020audeo} and general videos~\cite{di2021video,zhuo2023video}). Despite considerable advancements in the field of conditional symbolic music generation, the integration of diverse prompt types (such as images, videos, texts, tags, and humming) within a single generative model remains unexplored.

In this work, we aim to build a generalized, controllable and high-quality framework, referred to as \framework, for symbolic music generation. We address the challenges associated with this goal from four perspectives: \textbf{input}, \textbf{representation}, \textbf{assessment} and \textbf{data}, each described as follows: 

\textbf{1) Input: multi-modal parsing.} A versatile framework should support various multi-modal prompts as inputs. Given the inherent differences among multi-modal data, the primary challenge in multi-modal prompt parsing lies in effectively processing and extracting musical information from heterogeneous data sources. To address this, we propose a multi-modal prompt parsing method, termed \projector. This projector contains a novel projection space for symbolic music elements, serving as a bridge between diverse multi-modal prompts and core musical elements. In \projector, multiple prompt types (\textit{i.e.}, images, videos, texts, tags, and humming) are analyzed and mapped to specific musical elements (\textit{i.e.}, emotions, genres, rhythms, and notes), as shown in Fig.~\ref{fig:intro}. For instance, temporally-related prompts, such as videos and humming, are translated into rhythm elements to maintain temporal consistency. Emotional prompts, such as images, videos, and texts, are mapped to corresponding emotional elements to ensure the generated music accurately conveys the intended emotions. This approach harmonizes multi-modal prompts by translating them into a unified projection space of musical elements.

\textbf{2) Representation: precise control.} An optimal representation should contain fine-grained, type-specific musical elements (such as emotions, rhythms, 
and genres), facilitating accurate, efficient, and controllable music generation. In this paper, we build our music representation based on compound words~\cite{hsiao2021compound} with enhancements to musical elements. Specifically, the family tokens of our representation include note-related, rhythm-related, tag-related and 
instrument-related tokens. The tag-related tokens provide control over emotions and genres, while the instrument-related tokens (program) distinguish between different instrument tracks. With this representation, our music generator can efficiently produce coherent, melodious, and harmonious compositions aligned with the control signals generated by \projector.

\textbf{3) Assessment: high-quality music selection.} Prevailing methods generate final music outputs in a single pass, often resulting in inconsistent quality. Post-hoc music quality assessment is crucial yet overlooked in existing approaches. Automatically evaluating and selecting high-quality generated music is essential for ensuring superior outcomes. To this end, we propose a Selector that identifies high-quality music via a multi-task learning scheme, as shown in Fig.~\ref{fig:intro}. Recognizing that emotions, genres, and quality are global semantic concepts, we concurrently train the model on emotion recognition, genre recognition, and quality assessment tasks. This multi-task approach promotes knowledge transfer across these tasks, enhancing the model's ability to assess quality by leveraging insights from emotion and genre recognition. As a result, our approach achieves reliable quality assessment performance with a minimal number of annotated samples.

\textbf{4) Data: large-scale dataset.} High-quality, large-scale symbolic music datasets with fine-grained emotion and genre annotations are scarce and challenging to collect. To address this gap, we construct \dataset, a large-scale dataset comprising 108,023 MIDI files with precise and diverse emotion and genre labels. The \dataset dataset is approximately 10 times larger than the previous largest emotion-labeled dataset ELMG~\cite{bao2022generating} in terms of song size, as shown in Table~\ref{tab:dataset_comp}.

Our main contributions are summarized as follows:

\begin{itemize}
    \item We introduce a multi-modal controllable framework, termed \framework, for symbolic music generation. \framework supports various types of prompts (\textit{i.e.}, images, videos, texts, tags, and humming) as inputs and generates emotionally controllable, high-quality music tailored to the provided prompts.
    \item We propose \projector to parse various input prompts into specific symbolic music elements. These elements then serve as control signals that guide the music generation process.
    \item We design a music composer called \composer, which includes a Generator that creates music by following control signals, and a Selector that evaluates and filters the generated music via a multi-task learning scheme.
    \item We build \dataset, the largest symbolic music dataset to date. It is manually annotated by experts to facilitate automatic music generation with precise emotion and genre labels. The \dataset dataset will be made publicly available.
\end{itemize}

\section{Related Work}
\subsection{Artificial Intelligence Generated Content (AIGC)}
AIGC aims to utilize AI technology to automate content production while addressing human individual requirements. Recently, AIGC has demonstrated significant potential in generating high-quality content that closely resembles human-generated content (HGC), particularly in the areas of text generation~\cite{ouyang2022training,OpenAI2023GPT4TR,thoppilan2022lamda} and image synthesis~\cite{karras2019style,ramesh2022hierarchical,rombach2022high}. 
Despite recent progress, the field of music generation remains relatively underexplored within the AIGC community. AI-generated music still lacks the emotional depth and melodic richness typically found in human-composed music pieces. Recent studies~\cite{borsos2023audiolm,agostinelli2023musiclm,Forsgren_Martiros_2022,schneider2023mo,huang2023noise2music,copet2023simple} have focused on generating \textit{audio-based} music from textual inputs. However, \textit{audio-based} music generation faces challenges, such as limited editability and the inability to finely control attributes like tempo, pitch, duration, and rhythm. In contrast, \textit{symbolic} music, which represents musical ideas through notation, offers more flexible and precise control over these attributes. Therefore, this paper focuses on music generation in \textit{symbolic} domain. We present \framework, a universal symbolic music generation framework that supports flexible prompts, and \dataset, a large-scale symbolic music dataset annotated with precise emotion and genre labels.

\subsection{Symbolic Music Representations}

Conventional symbolic music representations can be classified into two main categories: image-like~\cite{Yang2017MidiNetAC,dong2018musegan} and MIDI-like~\cite{huangmusic,choi2020encoding,jiang2020transformer} representations. The image-like representation utilizes a 2D matrix, such as a binary piano roll~\cite{brunner2018symbolic,Dong2018ConvolutionalGA}, to indicate the presence of notes at each time position. Subsequently, convolutional operations are applied for music generation. In contrast, the MIDI-like representation encodes music as a sequence of events that evolve over time. Transformers~\cite{vaswani2017attention,dai2019transformer} are then utilized to capture the temporal dependencies between musical events. Nonetheless, MIDI-like representations possess inherent limitations in modeling music rhythm structure. REMI~\cite{huang2020pop} addresses this by organizing input data into a metrical structure, \textit{i.e.}, introducing positional elements such as bar and beat events, along with supportive musical information like tempo and chord events. Empirical evidence suggests that this approach improves the rhythmic regularity of the generated music. Compound Words~\cite{hsiao2021compound} further groups the REMI tokens by note type and metric type, significantly reducing token sequence length, thereby accelerating training and inference times. A recent study, SDMuse~\cite{sdmuse}, demonstrates the effectiveness of hybrid representations, leveraging the complementary strengths of image-like and MIDI-like representations for music editing and generation.

In this paper, we construct our music representation following the Compound Words structure with several crucial enhancements. We introduce two new family tokens: i) tag-related tokens (emotion, genre) to control emotional expression and musical style; and ii) instrument-related tokens (program) to facilitate the generation of multi-track music featuring diverse instruments.

\subsection{Symbolic Music Generation}

Given our interest in improving the versatility, controllability, and quality of symbolic music generation, it is crucial to review and compare various generation methods. We categorize these methods into five groups: unconditioned, tag-conditioned, sequence-conditioned, video-conditioned, and X-conditioned (our approach).

\subsubsection{Unconditioned Symbolic Music Generation} Unconditioned methods generate music from scratch using a random seed, without any specific constraints or additional input. The primary challenge lies in ensuring long-term structural coherence as the music length increases. Researchers focus on enhancing the overall repetitive structure of the generated music using various models, such as Transformer-based architectures~\cite{Dai2021ControllableDM,Zou2021MelonsGM}, RNN-based models~\cite{Medeot2018StructureNetIS,jhamtani2019modeling}, or optimization-based approaches~\cite{herremans2017morpheus}. In contrast, conditional music generation has gained popularity in recent years, as it enables users to guide the generation process to produce unique musical compositions.

\subsubsection{Tag-conditioned Symbolic Music Generation} Tag-conditioned methods involve conditioning the generation on high-level tags such as instrument, genre or emotion. For instance, MuseNet~\cite{payne2019musenet} can generate music based on a specific set of instruments and a particular musical style. GTR-CTRL~\cite{sarmento2023gtr} presents methods to condition Transformer-based models to generate guitar tabs based on the desired instrument and genre. EMOPIA~\cite{EMOPIA} is an emotion-labeled symbolic music dataset comprising 1,078 music clips from 387 songs, facilitating research on emotion-conditioned symbolic music generation~\cite{bao2022generating,neves2022generating,EmoMusicTV}.

\subsubsection{Sequence-conditioned Symbolic Music Generation} 
These methods typically employ a conditioning sequence as a prior to generate a continuation or extension accordingly. Standard sequence prompts for music generation include lead sheets, motifs, melodies, themes, and lyrics, which can be directly extracted from musical pieces to form training pairs. 
Huang \textit{et al.}~\cite{huangmusic} demonstrate the first successful application of Transformers to produce accompaniments conditioned on melodies. MELONS~\cite{Zou2021MelonsGM} is another transformer-based framework that generates full-song melodies with long-term structures given motifs. MGM~\cite{MGM2023} learns motif-level repetitions and integrates them into the music generation process. Theme Transformer~\cite{shih2022theme} introduces a theme-based conditioning approach that compels the model to manifest the given theme multiple times in its resulting generation. UP-Transformer~\cite{uptrans} focuses on user preference-based music transfer, utilizing a single piece of a user's favorite music as the condition for transferring musical styles. The relationship between melody and lyrics is crucial for symbolic music generation. Yu \textit{et al.}~\cite{yu2021conditional} propose a conditional LSTM-GAN model that generates melodies from lyrics by leveraging the syntactic structures of the lyrics through LSTM networks. This approach ensures the generated sequences align with the lyrics and mimic the distribution of real melody samples. Zhang \textit{et al.}~\cite{zhang2024controllable} present a novel Transformer-based approach to generate syllable-level lyrics from melodies, employing an explicit n-gram (EXPLING) loss function and a prior attention mechanism to improve sequence alignment and controllable lyric generation. Duan \textit{et al.}~\cite{duan2023melody} address the challenge of interpretability in melody generation from lyrics by integrating mutual information constraints and Transformer-based semantic feature extraction.

\subsubsection{Video-conditioned Symbolic Music Generation} \label{sec:video_conditiond_methods} Most previous methods for video-conditioned music generation focus on composing music from silent performance videos~\cite{gan2020foley,su2020audeo}, a process akin to visual music transcription. The instrument type and rhythm can be inferred from visual cues (\textit{e.g.}, performance venue, musician's actions, \textit{etc}), which limits music diversity to some extend. Recent methods~\cite{NEURIPS2021_f4e369c0,zhu2022quantized} take dance or human action videos as input, generating music pieces that plausibly match the corresponding visual input. However, these methods cannot be applied to general videos as they rely on additional keypoint annotations. CMT~\cite{di2021video} generates background music for general videos by establishing rule-based rhythmic video-music relationships. To mitigate potential style conflicts caused by this rule-based design, V-MusProd~\cite{zhuo2023video} introduces semantic-level correspondence. This approach decouples music generation into three progressive stages (chords, melody, and accompaniment) and extracts video-music relational features (semantic, color, motion) for guidance.

\subsubsection{X-conditioned Symbolic Music Generation} This paper introduces the X-conditioned framework, where X represents various types of prompts, such as images, videos, text, tags, and humming. Our \framework is a multi-modal controllable symbolic music generation framework designed to support versatile prompts.

\begin{figure*}[!t]
\begin{center}
\includegraphics[width=1.0\linewidth]{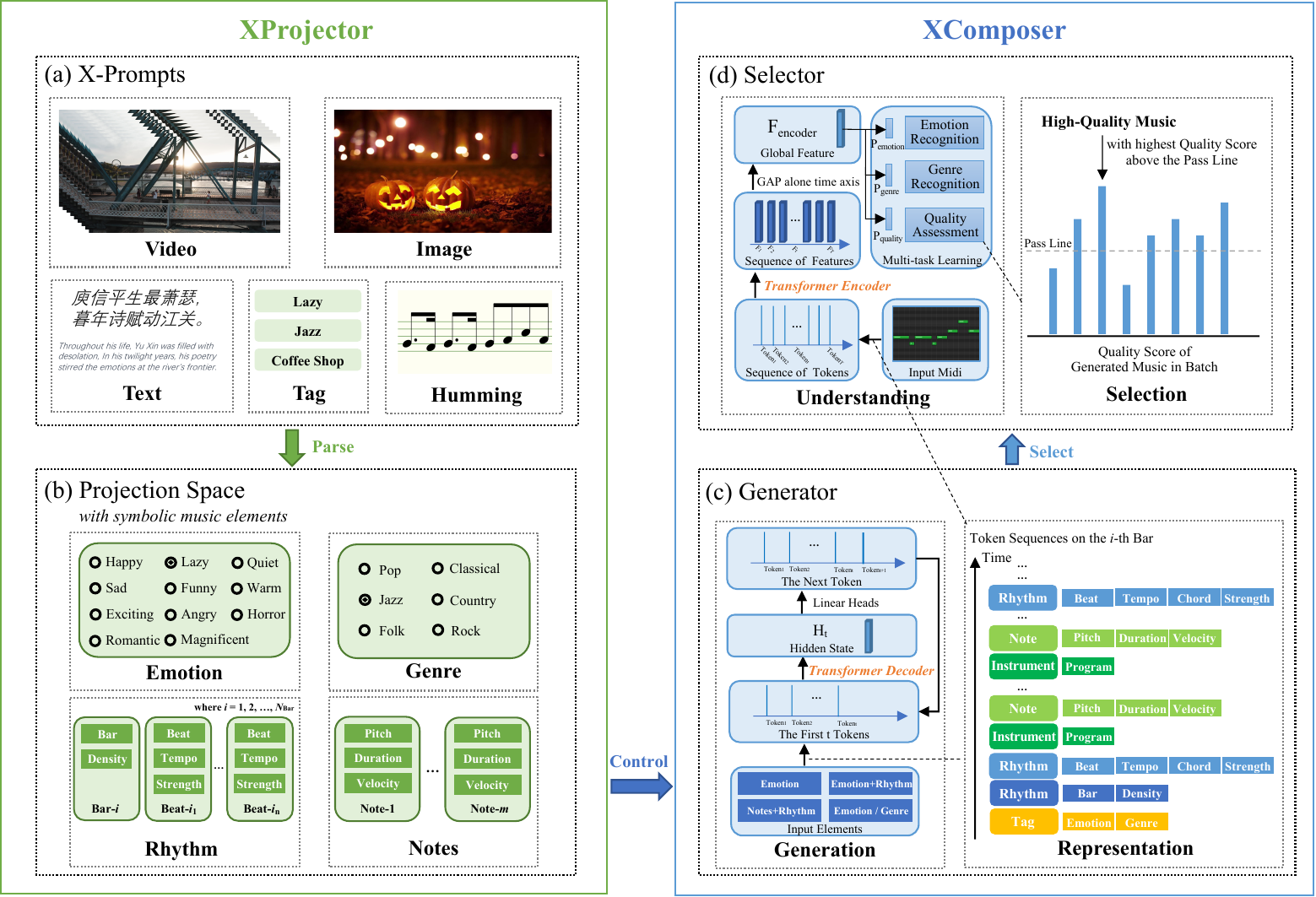}
\end{center}
   \caption{Illustration of the proposed \framework, which supports flexible (a) X-Prompts to guide the generation of high-quality symbolic music. The \projector analyzes these prompts, mapping them to symbolic music elements within the (b) Projection Space. Subsequently, the (c) Generator of \composer transforms these symbolic music elements into token sequences based on our enhanced representation. It employs a Transformer Decoder as the generative model to predict successive events iteratively, thereby creating complete musical compositions. Finally, the (d) Selector of \composer utilizes a Transformer Encoder to encode the complete token sequences and employs a multi-task learning scheme to evaluate the quality of the generated music.}
\label{fig:framework}
\end{figure*}

\section{Method}

Our proposed \framework supports various types of content as prompts for generating high-quality music. As shown in Fig.~\ref{fig:framework}, the process is divided into three stages: parsing, control, and selection. First, \projector (Sec.~\ref{sec:projector}) analyzes the input content and parses it into symbolic music elements within the projection space. Second, \composer (Sec.~\ref{sec:composer}) maps these elements to token sequences and controls the Generator to generate corresponding music. Finally, the Selector evaluates the quality of the generated music batches  and selects the one with the highest quality score. The symbolic music dataset \dataset is introduced in Sec.~\ref{sec:dataset}.

\subsection{\projector} \label{sec:projector}

\textbf{The projection space of symbolic music elements}, denoted as $\mathcal{P}$, acts as a bridge between multi-modal content and symbolic music. This space includes four types of symbolic music elements, emotions ($\mathcal{P}^E$), genres ($\mathcal{P}^G$), rhythms ($\mathcal{P}^R$), and notes ($\mathcal{P}^N$), represented as $\{\mathcal{P}^E, \mathcal{P}^G, \mathcal{P}^R, \mathcal{P}^N\} \in \mathcal{P}$. 

The emotion element $\mathcal{P}^E \in \mathbb{R}^{D_E}$ and the genre element $\mathcal{P}^G \in \mathbb{R}^{D_G}$ are expressed as one-hot vectors, where $D_E$ and $D_G$ denote the number of emotion and genre categories, respectively. In this paper, $\mathcal{P}^E$ can be chosen from 11 emotions: exciting, warm, happy, romantic, funny, sad, angry, lazy, quiet, fear, and magnificent. Similarly, $\mathcal{P}^G$ offers 6 common genre choices: rock, pop, country, jazz, classical, and folk.

The rhythm element spans the range of bars and is represented as $\mathcal{P}^R = \{ p^r_i \}_{i=1}^{N_{Bar}}$, where $N_{Bar}$ denotes the total number of bars. $p^r_i$ represents the rhythmic component of the $i$-\textit{th} bar and can be expanded as $\{p^{bar}_i, p^{beat}_{i_1},...,p^{beat}_{i_n}\}$, with $i_n$ indicating the number of beats within the bar. Specifically, the bar element $p^{bar} = ({\rm bar, density}) \in \mathbb{R}^2$ records the starting position and note density of the current bar, while the beat element $p^{beat} = ({\rm beat, tempo, strength}) \in \mathbb{R}^3$ captures the starting position, tempo and intensity of the beat. 

The note element covers the range of a single note and is defined as $\mathcal{P}^N=\{p^n_j\}^{N_{Note}}_{j=1}$, where $N_{Note}$ denotes the number of notes in the sequence. Each note element $p^n=({\rm pitch, duration, velocity}) \in \mathbb{R}^3$ represents the pitch, duration, and velocity of the note.

When prompts from various modalities are input, specific symbolic music elements are activated, guiding the matching music generation process. As shown in Table~\ref{tab:act_ele}, inputs such as videos, images, or text with emotional tendencies activate the emotion element. Similarly, inputs containing temporal information, like videos or humming, activate the rhythm element.

\begin{table}[!t]
\caption{Mapping table of the input prompts and the activated elements.}
\vspace{-10pt}
\label{tab:act_ele}
\begin{center}
\begin{tabular}{ccc}
\toprule
\textbf{Input Prompt} & \textbf{Analyzed Content} & \textbf{Activated Elements}\\
\midrule
video & emotion, rhythm & $\{\mathcal{P}^E, \mathcal{P}^R\}$ \\
image & emotion & $\{\mathcal{P}^E\}$ \\
text & emotion & $\{\mathcal{P}^E\}$ \\
tag & emotion / genre & $\{\mathcal{P}^E\} / \{\mathcal{P}^G\}$ \\
humming & notes, rhythm & $\{\mathcal{P}^R, \mathcal{P}^N\}$ \\
\bottomrule
\end{tabular}
\end{center}
\vspace{-15pt}
\end{table}

\textbf{\projector}, as the core component of \framework, analyzes multi-modal content and maps it into symbolic music elements within the projection space. The associated mapping function is denoted as $\mathcal{F}_{XP}$.

\textbf{Image prompts}, characterized by their non-sequential nature, guide the music generation process by controlling the overall properties of the sequence. Specifically, \projector performs sentiment analysis on the input image to determine its dominant emotion category and activates the corresponding emotion element within the projection space. This mechanism guides the generation of music aligned with the detected emotion. The image sentiment analysis module computes an emotion score $S^e({\rm image})$ for the input image and selects the emotion with the highest score. The calculation is as follows:

{\small
\begin{align}
\begin{split}
&S^e({\rm image}) = \lambda_1 * S^e_{\rm ResNet}({\rm image}) + \lambda_2 * S^e_{\rm CLIP}({\rm image}) \\
&\mathcal{F}_{XP}({\rm image}) = \{\mathcal{P}^E\} = \{\mathop{\rm argmax}_{e \in \mathcal{E}} S^e({\rm image})\}
\label{equation:image}
\end{split}
\end{align}
}

\noindent where $\mathcal{E}$ denotes the set of emotion categories. \projector employs two models for this calculation. The first model is the well-established deep convolutional neural network ResNet~\cite{he2016deep}. Specifically, we utilize the ResNet-50 architecture to train an image emotion classifier on our large-scale image emotion dataset (details in Sec.~\ref{sec:exp_dataset}). The classifier outputs the probability $S^e_{\rm ResNet}({\rm image})$ for each emotion $e$. The second model leverages CLIP~\cite{radford2021learning}, a prominent image-text pre-training model. We compute the embedding similarity between the input image and synonymous textual descriptions of each emotion $e$. These similarities are then normalized via the Softmax function to derive $S^e_{\rm CLIP}({\rm image})$. Weight factors $\lambda_1$ and $\lambda_2$ balance the contributions of the two models, with values set to $\lambda_1 = 1$ and $\lambda_2 = 2$ in our implementation.

\textbf{Text prompts}, inherently sequential data, are processed similarly to image-conditioned inputs. \projector performs sentiment analysis to identify the dominant emotion in the input text and activates the corresponding element within the projection space. The text sentiment analysis module employs the SentenceTransformer~\cite{reimers-2019-sentence-bert} model to calculate embedding similarities between the input text and synonymous descriptions of each emotion $e$. These similarities are then normalized via the Softmax function to produce an emotion score $S^e({\rm text})$ as follows:

{\small
\begin{equation}
\mathcal{F}_{XP}({\rm text}) = \{\mathcal{P}^E\} = \{\mathop{\rm argmax}_{e \in \mathcal{E}} S^e({\rm text})\}
\end{equation}
}

In \textbf{tag-conditioned} music generation, users can select from 11 emotion tags and 6 genre tags to guide the process. Once a tag is chosen, \projector activates the corresponding emotion or genre element within the projection space:

{\small
\begin{align}
\begin{split}
&\mathcal{F}_{XP}({\rm tag}^e) = \{\mathcal{P}^E\} = \{{\rm tag}^e\} \\
&\mathcal{F}_{XP}({\rm tag}^g) = \{\mathcal{P}^G\} = \{{\rm tag}^g\}
\end{split}
\end{align}
}

\textbf{Video prompts}, which are spatio-temporal data, guide both global and local attributes of the music sequence. For video-conditioned music generation, \projector analyzes sentiment, motion, and scene transitions within the input video, mapping these factors to the appropriate rhythm and emotion elements in the projection space. This ensures a high degree of synchronization between the generated music and the video content.

We observe a significant correlation between video background music tempo and scene transition frequency. For example, montage videos with rapid scene transitions typically feature fast-paced music, while peaceful scenery videos are often paired with slower tempos. To formalize this relationship, we introduce a scene transition rate metric $R_{\rm scene} $ to control the music tempo ${\rm t}_{\rm music}$:

{\small
\begin{align}
\begin{split}
&R_{\rm scene} = \frac{N_{\rm scene}}{T_{\rm video}}\\
&{\rm t}_{\rm music} = {\rm t}_{\rm init} + {\rm t}_{\rm inc} * tanh(R_{\rm scene})
\end{split}
\end{align}
}

Here, $N_{\rm scene}$ denotes the total number of scene transitions (computed using PySceneDetect~\cite{pyscenedetect}), while $T_{\rm video}$ represents the video duration in seconds. The music tempo ${\rm t}_{\rm music}$ (measured in bpm) is derived from $R_{\rm scene}$, with an initial tempo ${\rm t}_{\rm init}$ and incremental tempo ${\rm t}_{\rm inc}$. The $tanh$ activation function ensures that the coefficient for ${\rm t}_{\rm inc}$ remains between 0 and 1. 
Our analysis of tempo distributions in the training dataset shows that 98.6\% of musical tempos fall within the 60$\sim$130 bpm range. Accordingly, we set ${\rm t}_{\rm init}=60$ and ${\rm t}_{\rm inc}=70$ to keep generated tempos within this range.

Regarding emotions, we determine the emotional category of the input video through sentiment analysis and activate the corresponding emotion element within the projection space to control the emotional style of the music. The video sentiment analysis module computes an emotion score $S^e({\rm video})$ for the video and selects the emotion with the highest score as the analysis result. The calculation formula is as follows:

{\small
\begin{align}
\begin{split}
&S^e({\rm bar})=\frac{\sum_{m=1}^{N_{ipb}}S^e({\rm image}_m)}{N_{ipb}} \\
&N_{\rm bar} = \frac{T_{\rm video} * {\rm t}_{\rm music}}{60 * N_{bpb}} \\
&S^e({\rm video})=\frac{\sum_{i=1}^{N_{bar}}S^e({\rm bar}_i)}{N_{\rm bar}}
\end{split}
\end{align}
}

Here, we uniformly sample $N_{ipb}$ frames per bar and compute an emotion score $S^e({\rm image})$ for each frame. These scores are averaged to obtain a bar-level emotion score $S^e({\rm bar})$. Given $N_{bpb}$, the number of beats per bar, and using the video duration $T_{\rm video}$ along with the music tempo ${\rm t}_{\rm music}$, we calculate the total number of music bars $N_{\rm bar}$. Averaging $S^e({\rm bar})$ across all $N_{\rm bar}$ bars yields the final video emotion score $S^e({\rm video})$. In this paper, $N_{bpb}$ is set to 4, and $N_{ipb}$ is set to 8.

Inspired by CMT~\cite{di2021video}, which establishes a correlation between fast motion and dense notes, we control the local music rhythm using video motion information. Specifically, \projector extracts video motion information by calculating the optical flow ${\rm flow}_t(x, y)$. Due to the high computational complexity of optical flow, we use a more efficient PA~\cite{zhang2019pan} to model video motion. This average optical flow intensity within each bar, along with the visual beat saliency\cite{davis2018visual} for each beat, is mapped to note density and beat strength, respectively. Note that we also preserve their percentile distribution (based on the training set statistics)~\cite{di2021video} within the projection space. The calculation formula is:

{\small
\begin{align}
\begin{split}
&F_t = \frac{\sum_{x,y}|{\rm flow}_t(x, y)|}{HW}\\
&N_{fpb} = \frac{T_{\rm video} * {\rm fps}_{\rm video}}{N_{\rm bar}}\\
&{\rm density}_i \sim \frac{\sum_{t \in bar_i}F_t}{N_{fpb}}\\
&{\rm strength}_{i;j} \sim {\rm vbs}_{{\rm beat}_{i;j}}
\end{split}
\end{align}
}

Thus, the complete mapping relationship is expressed as follows:

{\small
\begin{align}
\begin{split}
&p^{\rm bar}_i = ({\rm bar}_i, {\rm density}_i) = (\frac{T_{\rm video} * (i - 1)}{N_{\rm bar}}, {\rm density}_i)\\
&p^{\rm beat}_{i;j} = ({\rm beat}_{i;j}, {\rm tempo}_{i;j}, {\rm strength}_{i;j})\\
&\quad\quad= ({\rm bar}_i + \frac{T_{\rm video} * (j - 1)}{N_{\rm bar} * N_{bpb}}, {\rm t}_{\rm music}, {\rm strength}_{i;j})\\
&\mathcal{F}_{XP}({\rm video}) = \{\mathcal{P}^E, \mathcal{P}^R\}\\
&\quad\quad\quad\quad\quad= \{\mathop{\rm argmax}_{e \in \mathcal{E}} S^e({\rm video}), \{p^{\rm bar}_i, \{p^{\rm beat}_{i;j}\}_{j=1}^{N_{bpb}}\}_{i=1}^{N_{\rm bar}} \}
\end{split}
\end{align}
}

\noindent where $i=1,2,...,N_{\rm bar}$ and $j = 1, 2, ..., N_{bpb}$.

\textbf{Humming prompts}, which are sequential data, guide the generation process of complete music by first being transcribed into an initial MIDI sequence. \projector employs the VOCANO~\cite{hsu2021vocano} algorithm to transcribe input humming audio into an original MIDI sequence, \textit{i.e.}, $M_{\rm origin} = {\rm VOCANO}({\rm humming})$. This sequence is then processed using standardization operations (beat processing and note quantization) to create a standard prior MIDI sequence, \textit{i.e.}, $M_{\rm std} = {\rm Standardize}(M_{\rm origin})$. For beat processing, the tempo of each beat is derived from the time intervals between adjacent beats, adjusting the default tempo of 120 bpm in the transcribed sequence to the actual tempo. Note quantization adjusts the onset and offset positions of each note to the nearest 32\textit{nd} note position. Finally, information from $M_{\rm std}$ is organized and mapped to the corresponding note and rhythm elements within the projection space to generate the subsequent music sequences. The detailed formulas are as follows:

{\small
\begin{align}
\begin{split}
&p^{\rm bar}_i = ({\rm bar}_i, {\rm density}_i)
= (T^{M_{\rm std}}_{\rm beat} * N_{bpb} * (i - 1), \mathcal{D}(M_{\rm std}^{{\rm bar}_i}))\\
&p^{\rm beat}_{i;j} = ({\rm beat}_{i;j}, {\rm tempo}_{i;j}, {\rm strength}_{i;j})\\
&\quad \quad= ({\rm bar}_i + T^{M_{\rm std}}_{\rm beat} * (j - 1), M_{\rm std}^{{\rm tempo}_{i;j}}, \mathcal{S}(M_{\rm std}^{{\rm beat}_{i;j}}))\\
&\mathcal{F}_{XP}({\rm humming}) = \{\mathcal{P}^N, \mathcal{P}^R\}\\
&\quad\quad\quad\quad\quad\quad\quad= \{\{M_{\rm std}^{{\rm note}_k}\}^{N_{\rm Note}}_{k=1}, \{p^{\rm bar}_i, \{p^{\rm beat}_{i;j}\}_{j=1}^{N_{bpb}}\}_{i=1}^{N_{\rm bar}}\}
\end{split}
\end{align}
}

\noindent where $i=1,2,...,N_{\rm bar}$, $j = 1, 2, ..., N_{bpb}$, and $\mathcal{D}$ and $\mathcal{S}$ denote the calculation formulas for note density and beat strength, respectively. $T^{M_{\rm std}}_{\rm beat}$ is the fixed beat length in $M_{\rm std}$, measured in seconds. $M_{\rm std}^{{\rm bar}_i}$, $M_{\rm std}^{{\rm beat}_{i;j}}$, and $M_{\rm std}^{{\rm tempo}_{i;j}}$ represent the $i$-th bar, the $j$-th beat within the $i$-th bar, and the tempo value for that beat in $M_{\rm std}$, respectively. These values contribute to calculating the rhythm elements.

\begin{figure}[!t]
\begin{center}
\includegraphics[width=1.0\linewidth]{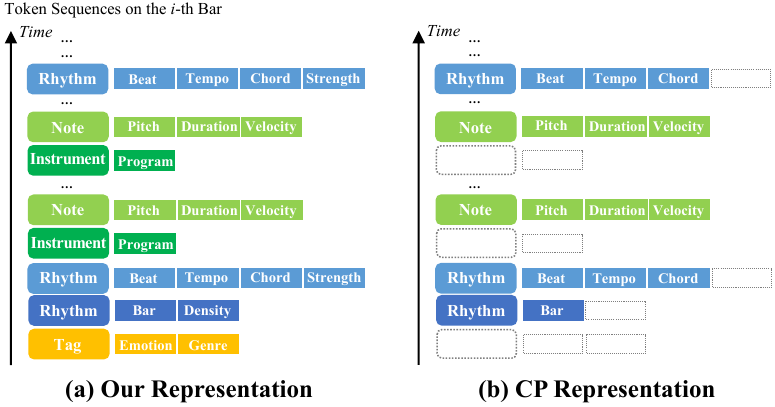}
\end{center}
   \caption{Comparison between our representation and Compound Word (CP) ~\cite{hsiao2021compound} representation. The dotted boxes represent our new tokens in comparison with those of the CP representation.}
\label{fig:representation}
\end{figure}

\subsection{\composer} \label{sec:composer}

\textbf{Our enhanced symbolic music representation}, as shown in Fig.~\ref{fig:representation}, maps MIDI files and the elements within the projection space to token sequences representing symbolic music, thereby assisting \composer in the subsequent music generation and selection processes.

\composer follows the Compound Word (CP)~\cite{hsiao2021compound} architecture, where tokens belonging to the same family (representing the same event) are grouped into a supertoken and positioned at the same time step. As illustrated in Fig.~\ref{fig:representation}, \composer introduces three key improvements:

\begin{itemize}
    \item First, we introduce a new family token named ``Tag'', along with two corresponding grouped tokens: ``Emotion'' and ``Genre''. These tokens enable control over the music generation process by specifying emotion and genre.
    \item Second, we add a new family token called ``Instrument'', with its corresponding grouped token, ``Program'', ensuring the generation of multi-track music.
    \item Third, within the ``Rhythm'' family token, we incorporate the grouped tokens ``Density'' and ``Strength'' into bar and beat events, respectively, allowing control over note density and beat strength.
\end{itemize}

In this paper, we utilize the Tag token to capture the overall semantic information of the music. Unlike methods that specify the Tag token solely at the beginning of the entire music piece (\textit{e.g.}, EMOPIA~\cite{EMOPIA}), our approach places the Tag token at the beginning of each bar. This strategy offers two key advantages: it generates music that adheres more closely to the specified tag and facilitates bar-by-bar fine-tuning of emotion categories in video-conditioned music generation scenarios. The Emotion and Genre tokens represent the emotional and stylistic characteristics of the music, offering 11 and 6 options, respectively. 

The Instrument token, positioned at the beginning of the note sequence, indicates the instrument information of the subsequent note sequence. This token enables track-level modeling for 17 instruments (detailed in Sec.~\ref{sec:dataset}), resulting in music enriched by diverse instrumental ensembles.

For video-conditioned music generation, the local rhythm of the generated music is synchronized with the motion information of the video. Inspired by CMT~\cite{di2021video}, we incorporate the Density and Strength tokens at bar and beat event positions, respectively, to control the note density and beat strength of each bar. This method ensures that the generated rhythm closely aligns with the video content.

Our representation chronologically encodes symbolic music events (\textit{e.g.}, Tag, Bar, Beat, Instrument, and Note) in each bar of MIDI files to form token sequences. This design supports the generation of emotionally expressive and melodically coherent music.

\begin{table}[!t]
\caption{Comparison of the implementation details between the CP and our proposed representation.}
\label{tab:representation}
\begin{center}
\def\arraystretch{1.1}
\begin{tabular}{lcccc}
\toprule
\multirow{2.3}{*}{\textbf{Token Type}} & \multicolumn{2}{c}{\textbf{Vocabulary Size}} & \multicolumn{2}{c}{\textbf{Embedding Size}}\\
\cmidrule(lr){2-3}
\cmidrule(lr){4-5}
& \textbf{CP}~\cite{hsiao2021compound} & \textbf{Ours} & \textbf{CP}~\cite{hsiao2021compound} & \textbf{Ours}\\
\midrule
$[$instrument$]$ & - & 17 (+1) & - & 64 \\
\hdashline
$[$tempo$]$ & 58 (+2) & 65 (+2) & 128 & 256\\
$[$position/bar$]$ & 17 (+1) & 33 (+1) & 64 & 256\\
$[$chord$]$ & 133 (+2) & 133 (+2) & 256 & 256\\
\hdashline
$[$pitch$]$ & 86 (+1) & 256 (+1) & 512 & 1024\\
$[$duration$]$ & 17 (+1) & 32 (+1) & 128 & 512\\
$[$velocity$]$ & 24 (+1) & 44 (+1) & 128 & 512\\
\hdashline
$[$family$]$ & 3 & 5 & 32 & 64\\
\hdashline
$[$density$]$ & - & 33 (+1) & - & 128\\
$[$strength$]$ & - & 37 (+1) & - & 128\\
\hdashline
$[$emotion$]$ & - & 11 (+1) & - & 64\\
$[$genre$]$ & - & 6 (+1) & - & 64\\
\hdashline
total & 338 (+8) & 672 (+13) & - & -\\
\bottomrule
\end{tabular}
\end{center}
\vspace{-10pt}
\end{table}

Implementation details of our representation are provided in Table~\ref{tab:representation}. For melodic instrument notes, we utilize 128 tokens to represent pitch, following the standard MIDI format. For percussion instrument notes, which lack pitch information, we employ 128 pseudo-pitch tokens to denote different percussion types. To reduce the vocabulary size, tempos are quantized into 65 values (ranging from 32 to 224), and velocities into 44 values (ranging from 40 to 126). We also add additional ``Tag'' and ``Instrument'' family tokens to represent emotion/genre and instrument information, respectively. To capture shorter note durations, we use a higher resolution (32\textit{nd} notes) instead of the 16\textit{th} notes used in CP. Furthermore, to accommodate large-scale data training, we have increased the embedding size for each token.

\textbf{The Generator}, serving as the core component of \composer, conditionally generates symbolic music based on our enhanced representation. It employs a Transformer Decoder~\cite{vaswani2017attention} as the backbone network to effectively model the dependencies among tokens.

Specifically, given the first $t$ tokens and aiming to predict the next token, the process is structured as follows. Initially, the token sequences are transformed into a two-dimensional event matrix, where each element ${\rm event}_i^j$ represents the $j$-th attribute of the $i$-th token. Each ${\rm event}_i$ contains 12 dimensions: family type, emotion, genre, bar-beat, tempo, chord, density, strength, program, pitch, duration, and velocity. Subsequently, at time $i$, a linear projection is applied to each one-hot vector ${\rm event}_i^j$, producing a dense vector ${\rm embed}_i^j$. These dense vectors are then concatenated to form the dense representation ${\rm concat}_i$ for the current token. Following this, linear projection and positional encoding are applied to ${\rm concat}_i$ to obtain the input feature ${\rm input}_i$ for the Transformer network at time $i$. The input features from the first $t$ tokens are fed into the Transformer network to compute the hidden state $H_t$ at the current time step. The next event ${\rm event}_{t+1}$ is then predicted by applying multiple linear projections to $H_t$. In line with the approach proposed by \cite{hsiao2021compound}, the next family type is predicted first, followed by the prediction of the other attributes based on $H_t$ and the one-hot vector of the predicted family type. Cross-entropy loss is utilized to optimize this prediction process. The overall procedure can be expressed as follows:

{\small
\begin{align}
\begin{split}
&{\rm embed}_i^j = {\rm Linear}_{\rm embed}^j({\rm Onehot}({\rm event}_i^j))\\
&{\rm concat}_i = {\rm Concat}(\{{\rm embed}_i^j\}_{j=1}^{12})\\
&{\rm input}_i = {\rm Linear}_{\rm input}({\rm PositionalEncoding}({\rm concat}_i))
\end{split}
\end{align}
}

\noindent for $i=1,...,t$ and $j=1,...,12$.

{\small
\begin{align}
\begin{split}
&H_t = {\rm TransformerDecoder}(\{{\rm input}_i\}_{i=1}^t)\\
&FT_t = {\rm Linear}_{\rm output}^1(H_t)\\
&E_t^j = {\rm Linear}_{\rm output}^j({\rm Concat}(H_t, {\rm Linear}_{\rm embed}^1(FT_t)))\\
&{\rm event}_{t+1}^{1} = {\rm argmax}(FT_t)\\
&{\rm event}_{t+1}^j = {\rm argmax}(E_t^j)
\end{split}
\end{align}
}

\noindent for $j=2,...,12$.

During inference, a stochastic temperature-controlled sampling method is employed to enhance the diversity of the generated tokens.

\textbf{The Selector}, as another core component of \composer, identifies high-quality symbolic music through a multi-task learning scheme. It leverages the Transformer Encoder~\cite{vaswani2017attention} as its backbone network to evaluate the quality of symbolic music.

Only a subset of the music generated by the Generator achieves human-level quality, which is characterized by beautiful and coherent melodies, distinct tune variations, and alternating rhythmic structures. Our objective is to accurately identify these high-quality pieces using supervised learning. To this end, we generate batches of symbolic music under diverse control conditions via the Generator. This is followed by manual annotations for each piece to determine whether it meets human-level standards. Then a classification model is trained using this annotated dataset to evaluate the quality of symbolic music accurately.

Specifically, we design a multi-task learning scheme comprising quality assessment, emotion recognition, and genre recognition tasks to select high-quality music. The Selector represents each MIDI file as a token sequence using our representation and translates this sequence into a corresponding event matrix. The same transformation is applied to each event vector within the matrix as in the Generator, yielding the input features for the Transformer network. Since the Selector analyzes entire MIDI files rather than predicting subsequent events, the $T$ input features from all moments are fed into the encoder, producing output features $F_i$ at each time step. We then apply Global Average Pooling along the temporal dimension to derive the global feature vector $F_{\rm encoder}$. This global representation is subsequently passed through three fully connected layers, with output activations normalized to produce class probabilities for each task. The process can be expressed as follows:

{\small
\begin{align}
\begin{split}
&{\rm embed'}_i^{j} = {\rm Linear}_{\rm embed'}^j({\rm Onehot}({\rm event}_i^j))\\
&{\rm concat'}_i = {\rm Concat}(\{{\rm embed'}_i^j\}_{j=1}^{12})\\
&{\rm input'}_i = {\rm Linear}_{\rm input'}({\rm PositionalEncoding}({\rm concat'}_i))
\end{split}
\end{align}
}

\noindent for $i=1,...,T$ and $j=1,...,12$.

{\small
\begin{align}
\begin{split}
&\{F_i\}_{i=1}^T = {\rm TransformerEncoder}(\{{\rm input'}_i\}_{i=1}^T)\\
&F_{\rm encoder} = {\rm GAP}_{\rm time}(\{F_i\}_{i=1}^T)\\
&{\rm Prob}_{\rm genre} = {\rm Softmax}({\rm FC}_{\rm genre}(F_{\rm encoder}))\\
&{\rm Prob}_{\rm emotion} = {\rm Softmax}({\rm FC}_{\rm emotion}(F_{\rm encoder}))\\
&{\rm Prob}_{\rm quality} = {\rm Softmax}({\rm FC}_{\rm quality}(F_{\rm encoder}))
\end{split}
\end{align}
}

The rationale for employing a multi-task learning scheme lies in the subtle differences in quality assessment criteria across various music types. By integrating emotion and genre recognition tasks, the network gains a more holistic understanding of the music, thereby enhancing its overall selection accuracy. Moreover, although the training dataset originates from the Generator, the Selector demonstrates remarkable generalizability, effectively assessing the quality, emotion, and genre of symbolic music from unknown sources.

During inference, the Selector employs ${\rm Prob}_{\rm quality}$ as the quality score. It identifies the sample within a batch that achieves the highest quality score surpassing a predefined threshold, thereby selecting the most promising high-quality music piece.

\begin{figure*}[!htbp]
\begin{center}
\includegraphics[width=1.0\linewidth]{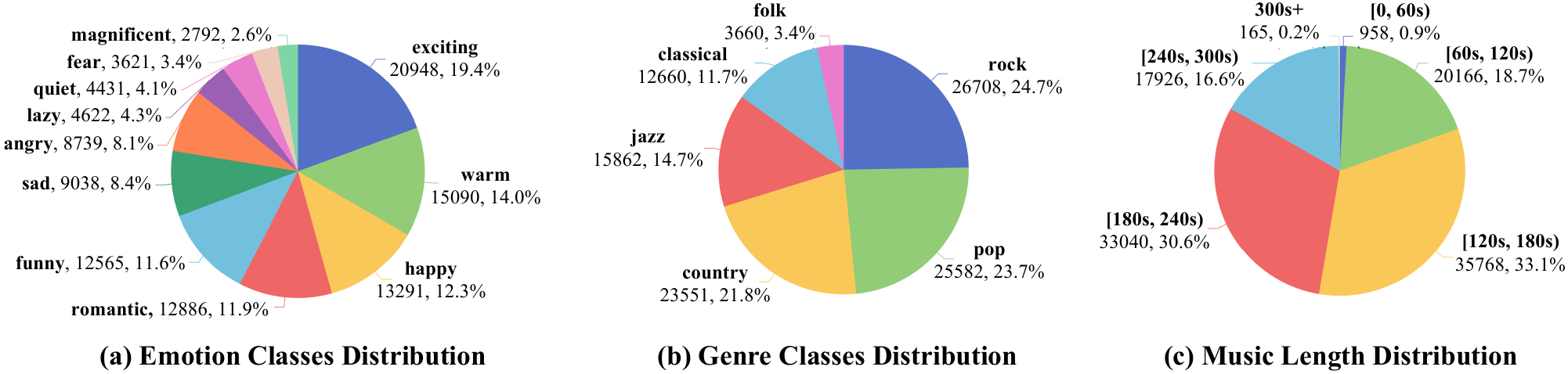}
\end{center}
   \caption{Data statistics of our \dataset dataset.}
\label{fig:dataset_stat}
\end{figure*}

\subsection{\dataset} \label{sec:dataset}

In this section, we introduce our constructed symbolic music dataset, \textit{i.e.}, \dataset. Existing publicly available symbolic music datasets suffer from limitations in both scale and label completeness, making it nearly impossible to train a music generation model that meets the requirements of this study. To address this gap, we built \dataset, the largest known symbolic music dataset with precise emotion and genre labels, comprising 108,023 MIDI files. The average duration of the music pieces is around 176 seconds, resulting in a total dataset length of around 5,278 hours.

For data collection and cleaning, we first crawled MIDI files from various online sources, including the Internet Archive, GitHub, and Reddit. To ensure dataset quality, we carried out the following data cleaning steps. i) \textit{Automatic Cleaning}: We removed corrupted or empty files and performed basic de-duplication based on MD5 file hashes. ii) \textit{Data Deduplication}: We rendered the remaining MIDI files into audio format, extracted chroma features, and calculated cosine similarities to identify and eliminate duplicates more effectively. iii) \textit{Manual Cleaning}: During the annotation phase, trained annotators discarded files with evident abnormalities (\textit{e.g.}, those with large missing note segments or excessively short sound effects), ensuring that only musically normal files remained. Following common practice~\cite{dong2018musegan}, we addressed data imbalance issue by merging instrument tracks. Specifically, we grouped the 128 melodic instruments into their parent categories. For example, program IDs 0$\sim$7 (such as Acoustic Piano and Electric Piano) were grouped under Piano, program IDs 24$\sim$31 (including Acoustic and Electric Guitars) under Guitar, and all 61 percussion instruments into the Drum category. This consolidation resulted in 17 distinct instrument types: piano, xylophone, organ, guitar, bass, violin, harp, string, trumpet, tuba, sax, flute, lead, pad, pipa, guzheng, and drum.

For data annotation, we established a comprehensive labeling system for emotions and genres and hired ten professional annotators (five males and five females) to ensure accurate labeling of each MIDI file. The annotators worked independently but maintained regular communication with the organizers to uphold consistent standards. To maintain high accuracy, we adopted several measures: i) \textit{Standardization}: Detailed descriptions and representative music demos were provided for each emotion and genre label to ensure a uniform understanding. ii) \textit{Cross-checking Mechanism}: Each annotation was independently verified by at least three experts. iii) \textit{Random Quality Checks}: The dataset was divided into batches of 500 files. Random samples were drawn from each batch for accuracy assessment, with batches failing to meet the 95\% accuracy threshold returned for revision. iv) \textit{Regular Training for the Annotators}: Weekly meetings were held to review frequently misclassified cases, reinforcing consistency among annotators. v) \textit{Discussion of Controversial Cases}: Controversial cases were deliberated upon by a panel consisting of all annotators and organizers.

\begin{table}[!t]
\caption{Comparison between existing emotion-labeled MIDI datasets and the proposed \dataset dataset.}
\label{tab:dataset_comp}
\begin{center}
\begin{tabular}{cccc}
\toprule
\textbf{Dataset} & \textbf{Emotion Type} & \textbf{Genre Type} & \textbf{Size (Songs)}\\
\midrule
MIREX-like~\cite{panda2013multi} & 5 classes & multiple & 193 \\
VGMIDI~\cite{ferreira-2019-learning} & valence & video game & 95 \\
EMOPIA~\cite{EMOPIA} & Russell's 4Q & pop & 387 \\
ELMG~\cite{bao2022generating} & 2 classes & multiple & 11,528 \\
\midrule
\textbf{XMIDI (Ours)} & 11 classes & 6 classes & \textbf{108,023} \\
\bottomrule
\end{tabular}
\end{center}
\vspace{-10pt}
\end{table}

The data statistics of our \dataset dataset are shown in Fig.~\ref{fig:dataset_stat}. In terms of emotion distribution (Fig.~\ref{fig:dataset_stat}a), categories such as exciting, warm, happy, romantic, funny, sad, and angry dominate, whereas emotion like lazy, quiet, fear, and magnificent are less frequent. The genre distribution (Fig.~\ref{fig:dataset_stat}b) is relatively balanced, with Rock representing the largest share and folk music the smallest. Regarding music length (Fig.~\ref{fig:dataset_stat}c), most compositions span between 1 and 5 minutes, with shorter and longer pieces being less common.

We compared \dataset dataset with existing emotion-labeled MIDI datasets in terms of emotion categories, genre types and data size. As shown in Table~\ref{tab:dataset_comp}, previous emotion datasets~\cite{panda2013multi,ferreira-2019-learning,EMOPIA} are relatively small, typically containing only a few hundred songs. Bao \textit{et al.}~\cite{bao2022generating} recently built a large-scale paired lyric-melody dataset annotated using deep learning models. However, their emotion labels are coarse-grained (positive/negative), and automated labeling introduces the risk of mis-classification. In contrast, our \dataset offers finer emotion categories (11 distinct emotions), more precise annotations (annotated by experts and cross-validated) and a nearly 10-times-larger data size.

\section{Experiments}

\subsection{Datasets} \label{sec:exp_dataset}

\subsubsection{Symbolic Music Generation} We developed XMIDI, the largest symbolic music dataset to date with precise emotion and genre labels. Each piece of music has an average duration of approximately 176 seconds, contributing to a cumulative dataset length of 5,278 hours.

\subsubsection{Image Emotion Recognition} We collected a new image emotion recognition dataset containing 269,793 images, among which 40,000 images are selected as the test set, and the remaining images are used for training. The images were gathered from various sources, including the WebEMO~\cite{panda2018contemplating} image sentiment dataset, the Places365~\cite{zhou2017places} scene recognition dataset, and web searches from Baidu and Google. Given that the labels of internet images are inherently noisy, we employed human annotators to filter out samples that did not clearly correspond to their assigned labels.

\subsubsection{Text Emotion Recognition} We constructed a standard test set for text emotion classification, consisting of 1,100 sentences distributed evenly across 11 emotion categories. To generate emotionally nuanced text, we instructed GPT-4~\cite{OpenAI2023GPT4TR} to produce sentences that conveyed specific emotions without explicitly including emotion words. We applied SentenceTransformer~\cite{reimers-2019-sentence-bert} to compute text embedding similarities and removed redundant samples to ensure diversity and distinctiveness within the dataset.

\subsubsection{Music Quality Assessment} We utilized the Generator of \composer to generate 9,540 music pieces by applying random combinations of emotion and genre tags as conditions. Subsequently, we conducted a manual quality evaluation of the generated pieces. High-quality music was characterized by coherent melodies, distinct tune fluctuations, and dynamic rhythmic variations. In contrast, pieces failing to meet these criteria did not achieve human-level quality standards.

\subsection{Implementation Details}

We employ the Transformer architecture~\cite{vaswani2017attention} as the backbone of \composer. For the Generator, we utilize Transformer Decoder to predict subsequent symbolic music event based on the previous events. The model consists of 30 self-attention layers, each containing 16 attention heads, with a hidden size set to 1,024. During training, the Adam optimizer is employed with an initial learning rate of 3e-5. When the loss saturates (specifically at values of 0.053, 0.049, and 0.045), we halve the learning rate and resume training. The overall training procedure spans 24 days, using a batch size of 40 across 8 NVIDIA A800 GPUs. To mitigate gradient explosion, we set the maximum gradient to 0.5. For the Selector, we use Transformer Encoder to encode the token sequences and predict global attributes such as quality levels, emotions, and genres. We use 3 self-attention layers, 8 attention heads, and a hidden size of 512. During training, the Adam optimizer is employed with a learning rate of 1e-5, and the model training process is completed in 10 hours on a single NVIDIA Tesla V100 GPU.

\begin{table}[!t]
\caption{Objective comparison with state-of-the-art symbolic music generation methods.}
\label{tab:obj_eval}
\begin{center}
\begin{tabular}{llccc}
\toprule
& \textbf{Method} & \textbf{PCE $\downarrow$} & \textbf{GS $\uparrow$} & \textbf{EBR $\downarrow$}\\
\midrule
\multirow{3}{*}{(a) Unconditioned} & CP~\cite{hsiao2021compound} & 2.6025 & 0.9990 & 0.0273 \\
& EMOPIA~\cite{EMOPIA} & 2.6756 & 0.9989 & 0.1197 \\
& \textbf{\framework (Ours)} & \textbf{2.5174} & \textbf{0.9992} & \textbf{0.0045} \\
\midrule
\multirow{2}{*}{(b) Video-conditioned} & CMT~\cite{di2021video} & 2.7290 & 0.6698 & 0.0321 \\
& \textbf{\framework (Ours)} & \textbf{2.6161} & \textbf{0.9983} & \textbf{0.0078} \\
\bottomrule
\end{tabular}
\end{center}
$^{\dagger}$PCE: Pitch Class Histogram Entropy; GS: Grooving Pattern Similarity; EBR: Empty Beat Rate.
\end{table}

\subsection{Objective Evaluation}

\subsubsection{Metrics} We selected three typical objective metrics: Pitch Class Histogram Entropy (PCE)~\cite{Wu2020TheJT}, Grooving Pattern Similarity (GS)~\cite{Wu2020TheJT} and Empty Beat Rate (EBR)~\cite{dong2018pypianoroll}. Specifically, the PCE evaluates the distribution and uniformity of pitch classes within a musical piece or segment. A lower PCE indicates a more concentrated pitch class distribution, usually implying clearer tonality. Conversely, a higher PCE represents a more uniform distribution, reflecting unstable tonality. The GS metric assesses the resemblance between rhythmic patterns in musical bars. A high GS score means that the grooving patterns of the analyzed pairs are similar, indicating a strong rhythm structure. In contrast, a low GS score means dissimilar grooving patterns, indicating a unstable rhythm structure. In addition, the EBR measures the proportion of empty or silent beats in a music piece. A higher EBR score suggests frequent gaps or pauses, indicating sparse note distribution and a lack of richness in the music. All three objective metrics are computed using the MusPy Toolkit~\cite{dong2020muspy}.

\subsubsection{Comparison with Symbolic Music Generation Methods} We compared XMusic with the current state-of-the-art symbolic music generation methods: CP~\cite{hsiao2021compound} and EMOPIA~\cite{EMOPIA}. For fair comparison, we used their official pre-trained models directly for inference. The comparative results are listed in Table~\ref{tab:obj_eval}-(a). Our XMusic outperforms both methods, achieving the lowest PCE and EBR scores and the highest GS score. This suggests that our method exhibits clear tonality, rich note distribution and distinct rhythmicity, respectively.

\subsubsection{Comparison with Video-conditioned Symbolic Music Generation Method} To further evaluate our method, we compared XMusic with CMT~\cite{di2021video}, the state-of-the-art video-conditioned symbolic music generation method. CMT is the first and only open-source method capable of generating background music for general videos. We did not include V-MusProd~\cite{zhuo2023video} in this comparison because it had not been open-sourced at the submission time of this manuscript. We randomly selected videos covering various scenes, including landscapes, animations, weddings, performances, sports, and games. These videos varied in duration (from 30 seconds to 2 minutes) and conveyed diverse sentiments, such as exciting, romantic, fear, happy emotions, \textit{etc}. We generated background music for each video using both CMT and our XMusic. The results are summarized in Table~\ref{tab:obj_eval}-(b). Compared with CMT, XMusic has clear advantages across all three objective metrics, demonstrating the effectiveness of our method.

\begin{table*}[!t]
\caption{Subjective comparison with state-of-the-art symbolic music generation methods.}
\vspace{-5pt}
\label{tab:subj_eval}
\begin{center}
\resizebox{1.0\linewidth}{!}{
\begin{tabular}{llcccccc}
\toprule
& \textbf{Method} & \textbf{Richness} & \textbf{Correctness} & \textbf{Structuredness} & \textbf{Emotion-Matching} & \textbf{Rhythm-Matching} & \textbf{Overall Rank}\\
\midrule
\multirow{3}{*}{(a) Unconditioned} & CP~\cite{hsiao2021compound} & 2.2323 & 2.1161 & 2.0871 & - & - & 2.1452 \\
& EMOPIA~\cite{EMOPIA} & 2.2387 & 2.4807 & 2.3452 & - & - & 2.3549 \\
& \textbf{\framework (Ours)} & \textbf{1.5290} & \textbf{1.4032} & \textbf{1.5677} & - & - & \textbf{1.5000} \\
\midrule
\multirow{2}{*}{(b) Video-conditioned} & CMT~\cite{di2021video} & 1.7129 & 1.7484 & 1.6742 & 1.6452 & 1.6807 & 1.6923 \\
& \textbf{\framework (Ours)} & \textbf{1.2871} & \textbf{1.2516} & \textbf{1.3258} & \textbf{1.3548} & \textbf{1.3194} & \textbf{1.3077} \\
\midrule
\multirow{3}{*}{(c) Text-conditioned} & BART-base~\cite{wu2022exploring} & 2.3871 & 2.4806 & 2.2226 & 2.6258 & - & 2.4290 \\
& GPT-4~\cite{OpenAI2023GPT4TR} & 2.1355 & 2.0065 & 2.1903 & 1.8613 & - & 2.0484 \\
& \textbf{\framework (Ours)} & \textbf{1.4774} & \textbf{1.5129} & \textbf{1.5871} & \textbf{1.5129} & - & \textbf{1.5226} \\
\midrule
\multirow{2}{*}{(d) Image-conditioned} & Synesthesia~\cite{tan2020automated} & 1.5548 & 1.8065 & 1.7484 & 1.7936 & - & 1.7258 \\
& \textbf{\framework (Ours)} & \textbf{1.4452} & \textbf{1.1935} & \textbf{1.2516} & \textbf{1.2064} & - & \textbf{1.2742} \\
\bottomrule
\end{tabular}
}

\end{center}
$^{\dagger}$The numbers listed in each block represent the average rankings of the comparison methods.
\end{table*}

\begin{table}[!t]
\caption{Subjective comparison with state-of-the-art emotion-conditioned symbolic music generation method.}
\vspace{-5pt}
\label{tab:subj_emo_music}
\begin{center}
\begin{tabular}{ccc}
\toprule
\textbf{Classes} & \textbf{EMOPIA~\cite{EMOPIA}} & \textbf{\framework (Ours)}\\
\midrule
Positive Valence (PV) $\uparrow$ & 38\% & \textbf{76\%} \\
Negative Valence (NV) $\uparrow$ & 38\% & \textbf{70\%} \\
Positive Arousal (PA) $\uparrow$ & 39\% & \textbf{66\%} \\
Negative Arousal (NA) $\uparrow$ & 51\% & \textbf{84\%} \\
\bottomrule
\end{tabular}
\vspace{-10pt}
\end{center}
\end{table}

\subsection{Subjective Evaluation}

For music generation, subjective human evaluations are essential, as they offer a comprehensive understanding of music quality. We designed online questionnaires for subjective evaluation and invited 31 participants to participate. To ensure blind evaluation, the music results within each question were randomly shuffled. 

\subsubsection{Comparison with Symbolic Music Generation Methods} Following common practice~\cite{hsiao2021compound,EMOPIA}, we generated MIDI files for each method (CP~\cite{hsiao2021compound}, EMOPIA~\cite{EMOPIA} and our proposed XMusic) and rendered these files in audio format using the same soundfont. 
In the questionnaire, each question contained three randomly ordered audio samples generated via the three aforementioned methods. The participants were required to carefully listen to and rank the audio samples based on the following metrics: i) \textit{Richness}: The diversity of musical elements, such as melody, harmony, rhythm, and timbre. ii) \textit{Correctness}: The absence of errors or unnatural musical elements, such as odd chords or sudden silences. iii) \textit{Structuredness}: The presence of repetitive structures, such as memorable melodies. The questionnaire took an average of 47 minutes to complete. We averaged the ranking results from the 31 participants to obtain the final results, presented in Table~\ref{tab:subj_eval}-(a). As shown, our method achieved the highest average rank across all three evaluation metrics, indicating that XMusic surpassed the existing state-of-the-art approaches in generating impressive and high-quality music.

\subsubsection{Comparison with Emotion-conditioned Symbolic Music Generation Method} \label{sec:emo_subj} To evaluate the efficacy of our method in emotion control, we compared our XMusic with the state-of-the-art emotion-conditioned method EMOPIA~\cite{EMOPIA}. To our best knowledge, EMOPIA is currently the only open-source emotion-conditioned symbolic music generation method. EMOPIA and XMusic utilize differing levels of emotion granularity. EMOPIA adopts Russell's Circumplex model, conceptualizing emotions in a two-dimensional space defined by valence and arousal, resulting in four classes (quadrants): PVPA (positive valence positive arousal), NVPA (negative valence positive arousal), NVNA (negative valence negative arousal), and PVNA (positive valence negative arousal). In contrast, XMusic employs 11 specific emotion classes, including happy, funny, sad, exciting, \textit{etc}. Since there are no clear correspondences between the 4 EMOPIA quadrants and the 11 XMusic classes, we aligned the emotion categories by merging adjacent EMOPIA quadrants and mapping corresponding XMusic categories to these combined classes.
For example, PVPA and PVNA were merged to form a new Positive Valence (PV) class, with the ``happy'' and ``funny'' classes mapped to this category. The other new categories were defined as follows: NV (combining NVPA \& NVNA, mapped to sad), PA (PVPA \& NVPA, mapped to exciting) and NA (PVNA \& NVNA, mapped to quiet). We generated and rendered music files for each method using these new categories as input prompts. Participants were instructed to count the number of music pieces that they perceived as fitting the new labels. This task took an average of 112 minutes to complete. We computed the average number of correctly classified pieces per class as determined by the 31 participants. As shown in Table~\ref{tab:subj_emo_music}, the music generated by our method better matched the input emotion prompts, demonstrating that our approach has superior emotional controllability.

\subsubsection{Comparison with Video-conditioned Symbolic Music Generation Method} Additionally, we compared XMusic with CMT~\cite{di2021video} in a video-conditioned evaluation. While the latest work, V-MusProd~\cite{zhuo2023video}, also focuses on generating music for general videos, the source code for V-MusProd was not publicly available at the time of our evaluation. Therefore, a direct and fair comparison could only be conducted between XMusic and CMT. In addition to evaluating richness, correctness, and structuredness, we assessed the degree of video-music alignment, focusing on both emotional and rhythmic correspondences. Using the same videos selected for the objective evaluation, we paired each video with background music generated by CMT and XMusic. These were presented in a random order for blindness. The questionnaire took about 25 minutes to complete. The participants were required to carefully listen and rank the two background music pieces in terms of five aspects. The average rankings from the 31 participants are summarized in Table~\ref{tab:subj_eval}-(b). XMusic consistently outperformed CMT across all five metrics, demonstrating its superior performance in handling video prompts. We attribute this success to the powerful emotion analysis and control capabilities of XMusic. In contrast, CMT relies on rule-based rhythm control, which lacks perception and control of emotions, resulting in poor emotional alignment. For instance, it may even generate cheerful music for a sorrowful video. By effectively understanding and controlling emotions, XMusic processes semantic information to generate music that harmonizes both rhythmically and emotionally. Video demos are available on our \MYhref{https://sites.google.com/view/xmusicdemos}{demo website}.

\subsubsection{Comparison with Text-conditioned Symbolic Music Generation Methods} We compared XMusic with two existing text-conditioned methods: BART-base~\cite{wu2022exploring} and GPT-4~\cite{OpenAI2023GPT4TR} (instructed to produce ABC notation, following~\cite{bubeck2023sparks}). For fair comparison, the ABC notation outputs were converted to MIDI format and rendered using the same soundfont. Participants evaluated music generated from identical text inputs via a structured questionnaire, ranking the results based on four evaluation metrics. The average questionnaire completion time was 34 minutes. As listed in Table~\ref{tab:subj_eval}-(c), \framework outperformed both comparative methods across all metrics, demonstrating that the idea of explicitly analyzing emotions in text and using this analysis to control music generation is effective.

\subsubsection{Comparison with Image-conditioned Symbolic Music Generation Methods} Symbolic music generation using images is relatively under-explored, with only a few notable methods~\cite{tan2020automated,wu2008study,madhok2018sentimozart} aiming to discover visual-musical associations. Among these, we could only compare XMusic with Synesthesia~\cite{tan2020automated}, as the source codes and demos for the other methods were unavailable. Specifically, we used the same images provided in the official Synesthesia repository~\footnote[2]{\url{https://github.com/sudongtan/synesthesia}} as inputs to our XMusic and then created a questionnaire to rank the two methods given the same input image. The questionnaire took about 13 minutes to complete. Table~\ref{tab:subj_eval}-(d) shows the average rankings from 31 participants. In contrast to Synesthesia, which implicitly models emotional information using paired image-music data, our \framework explicitly decouples the emotion analysis and control processes, offering a more intuitive and effective solution.

\begin{table*}[!t]
\caption{Ablation analysis.}
\vspace{-5pt}
\label{tab:abla_ana}
\begin{center}
\resizebox{1.0\linewidth}{!}{
\begin{tabular}{llcccccc}
\toprule
& \textbf{Setting} & \textbf{Richness} & \textbf{Correctness} & \textbf{Structuredness} & \textbf{Emotion-Matching} & \textbf{Rhythm-Matching} & \textbf{Overall Rank}\\
\midrule
\multirow{2}{*}{(a) Selector} & without (\xmark) Selector & 1.5957 & 1.5878 & 1.5484 & 1.5348 & - & 1.5667 \\
& with (\checkmark) Selector & \textbf{1.4043} & \textbf{1.4122} & \textbf{1.4516} & \textbf{1.4652} & - & \textbf{1.4333} \\
\midrule
\multirow{3}{*}{(b) Emotion Control} & No control & 2.2097 & 2.2903 & 2.2000 & 2.3548 & 2.3000 & 2.2710 \\
& Music-level & 2.0839 & 2.0065 & 2.2000 & 2.1129 & 2.1484 & 2.1103 \\
& Bar-level & \textbf{1.7064} & \textbf{1.7032} & \textbf{1.6000} & \textbf{1.5323} & \textbf{1.5516} & \textbf{1.6187} \\
\midrule
\multirow{4}{*}{(c) Music Representation} & CP~\cite{hsiao2021compound} & 3.2323 & 3.2942 & 3.7844 & - & - & 3.4370 \\
& CP+Tag & 2.8179 & 2.9640 & 2.5991 & - & - & 2.7937 \\
& CP+Tag+Instr & 2.4308 & 2.1195 & 2.1022 & - & - & 2.2175 \\
& CP+Tag+Instr+Rhythm (Ours) & \textbf{1.5190} & \textbf{1.6223} & \textbf{1.5143} & - & - & \textbf{1.5519} \\
\midrule
\multirow{3}{*}{(d) Data Sampling} & Undersampling & 2.3813 & 2.4746 & 2.1271 & - & - & 2.3277 \\
& Oversampling & 2.2069 & 2.0604 & 2.3667 & - & - & 2.2113 \\
& Original XMIDI & \textbf{1.4118} & \textbf{1.4650} & \textbf{1.5062} & - & - & \textbf{1.4610} \\
\bottomrule
\end{tabular}
}
\end{center}
\end{table*}

\subsubsection{Evaluation of the Controllability of Humming to Generate Music} Although some Apps, such as HumBeatz and ZhiQu, offer the capability to generate accompaniment based on user humming, few research works have focused on generating melodies from humming input. A recent work, Humming2Music~\cite{qiu2023humming2music}, is most relevant to the humming controllability of XMusic. However, a direct comparison was not feasible because the authors did not release their open-source code or demonstrations. Through a subjective evaluation following \cite{qiu2023humming2music}, we observed that XMusic has the following advantages: i) Accurate transcription. The transcription results were well-aligned with the original input humming melody. ii) Smooth transition. The transitions between transcribed humming notes and the subsequent composition were natural,  benefiting from the long-term dependencies captured by our \composer. iii) Consistent rhythm. The overall rhythm remained coherent, with no noticeable interruptions or abrupt changes. This consistency supports the effectiveness of our approach in parsing user humming and mapping it to notes and rhythmic elements.

\subsubsection{Evaluations on Public Datasets} We conducted experiments to compare XMusic with CP~\cite{hsiao2021compound} and EMOPIA~\cite{EMOPIA} on two widely used symbolic music datasets: \textit{AILabs1k7}~\cite{hsiao2021compound} and \textit{EMOPIA}~\cite{EMOPIA}. i) \textit{AILabs1k7} contains 1,748 pop piano MIDI files, each with an average duration of 4 minutes, totaling around 108 hours. Since this dataset lacks emotion and genre annotations, the emotion token in the EMOPIA model and the emotion and genre tokens in XMusic were set to [ignore]. ii) \textit{EMOPIA} consists of 1,087 MIDI clips extracted from 387 popular piano music pieces and includes emotion labels at the clip level. As genre labels are absent in this dataset, the genre token in the XMusic model was also set to [ignore]. Following \cite{EMOPIA}, we first pre-trained our model on the AILabs1k7 dataset due to the relatively small scale of EMOPIA dataset. The subjective evaluation results are listed in Table~\ref{tab:public_dataset}. With identical training data, our XMusic outperformed both CP and EMOPIA, demonstrating the effectiveness and generalizability of the proposed method. As discussed in Sec.~\ref{sec:effect_selector} and Sec.~\ref{sec:effect_rep}, the performance gains are primarily attributed to the enhanced music representation and the effective Selector model.

\begin{table}[!t]
\caption{Comparison with state-of-the-art symbolic music generation methods on public datasets.}
\vspace{-5pt}
\label{tab:public_dataset}
\begin{center}
\resizebox{1.0\linewidth}{!}{
\begin{tabular}{llcccccc}
\toprule
\textbf{Dataset} & \textbf{Method} & \textbf{Richness} & \textbf{Correctness} & \textbf{Structuredness} & \textbf{Overall Rank}\\
\midrule
\multirow{3}{*}{\textit{AILabs1k7}~\cite{hsiao2021compound}} & CP~\cite{hsiao2021compound} & 2.1521 & 2.1053 & 2.0276 & 2.0950\\
& EMOPIA~\cite{EMOPIA} & 2.2567 & 2.2318 & 2.2382 & 2.2422\\
& \textbf{XMusic (Ours)} & \textbf{1.5912} & \textbf{1.6629} & \textbf{1.7342} & \textbf{1.6628}\\
\midrule
\multirow{2}{*}{\textit{AILabs1k7}~\cite{hsiao2021compound}} & CP~\cite{hsiao2021compound} & 2.1016 & 2.0651 & 2.0177 & 2.0615\\
\multirow{2.5}{*}{+\textit{EMOPIA}~\cite{EMOPIA}} & EMOPIA~\cite{EMOPIA} & 2.2607 & 2.1889 & 2.2956 & 2.2484\\
& \textbf{XMusic (Ours)} & \textbf{1.6377} & \textbf{1.7460} & \textbf{1.6867} & \textbf{1.6901}\\
\bottomrule
\end{tabular}
}
\vspace{-15pt}
\end{center}
\end{table}

\subsection{Ablation Study}

\subsubsection{The Effectiveness of the Selector}\label{sec:effect_selector} To validate the effectiveness of our quality assessment model, we investigated whether incorporating the Selector improved music quality. Specifically, we employed two models, one with the Selector and one without, to generate 5 music pieces for each emotion. We then randomly selected two pieces belonging to the same emotion but originating from different models to create comparative pairs, yielding a total of 55 pairs. The participants were required to rank the music pieces in each pair from 4 perspectives: richness, correctness, structuredness and emotional matching. The completion of this questionnaire took an average of 143 minutes. The average rankings for the two models are presented in Table~\ref{tab:abla_ana}-(a). As shown, the music generated using the Selector consistently received higher average rankings across all four metrics, demonstrating the effectiveness of our Selector model and the necessity of conducting post-hoc music quality assessments.

We also conducted an objective ablation study on the Selector. As shown in Table~\ref{tab:obj_abla_selector}, incorporating the Selector improved the GS score and reduced the PCE and EBR scores, objectively demonstrating the effectiveness of the proposed Selector. In summary, the effectiveness of the Selector has been validated through both subjective and objective evaluations.

As described in Sec.~\ref{sec:composer}, our Selector leverages a multi-task learning scheme involving three sub-tasks: music quality assessment, emotion recognition, and genre recognition. We conducted ablation experiments to validate the effectiveness of the multi-task learning scheme. As shown in Table~\ref{tab:multitask_abla_selector}, the accuracy of the music quality assessment task significantly improved as when additional sub-tasks were incrementally added in joint learning. We conjecture that this improvement stemmed from the fact that music quality, emotion, and genre are subjective attributes influenced by human perception. Joint learning across these tasks helps the Selector align more closely with human perceptual standards. The Selector achieved a music quality assessment accuracy of 94.8\% on our self-constructed evaluation benchmark, demonstrating its robust quality control capabilities.

In summary, subjective evaluations, objective results, and multi-head ablation studies collectively demonstrate the effectiveness of our Selector.

\begin{table}[!t]
\caption{Objective ablation study on the proposed Selector.}
\label{tab:obj_abla_selector}
\begin{center}
\begin{tabular}{lccc}
\toprule
\textbf{Setting} & \textbf{PCE $\downarrow$} & \textbf{GS $\uparrow$} & \textbf{EBR $\downarrow$}\\
\midrule
w/o (\xmark) Selector & 2.5806 & 0.9991 & 0.0097\\
with (\checkmark) Selector & 2.5083 & 0.9994 & 0.0042 \\
\bottomrule
\end{tabular}
\end{center}
\end{table}

\begin{table}[!t]
\caption{Ablation study on the multi-task learning scheme of the Selector.}
\label{tab:multitask_abla_selector}
\begin{center}
\begin{tabular}{cccc}
\toprule
\multicolumn{3}{c}{\textbf{Classification Head}} & \multirow{2.3}{*}{\textbf{Accuracy}} \\
\cmidrule(lr){1-3}
\textbf{Quality} & \textbf{Emotion} & \textbf{Genre}\\
\midrule
\checkmark & \xmark & \xmark & 83.2\%\\
\checkmark & \checkmark & \xmark & 90.1\% \\
\checkmark & \checkmark & \checkmark & \textbf{94.8\%} \\
\bottomrule
\end{tabular}
\end{center}
\end{table}

\subsubsection{The Effectiveness of Fine-Grained Emotion Control} As mentioned in Sec.~\ref{sec:composer}, we introduce an emotion token before each bar to enable fine-grained emotion fine-tuning at the bar level. To verify the impact of this enhancement on video-conditioned music generation, we designed three comparative settings: i) \textit{no control}, \textit{i.e.}, without specifying the emotion of the generated music; ii) coarse-grained \textit{music-level} control, \textit{i.e.}, using only the video-level emotion tag as the initial emotion token to generate music; and iii) fine-grained \textit{bar-level} control, \textit{i.e.}, using the video-level emotion tag as the initial emotion token and fine-tuning with bar-specific emotion tags during inference. The participants ranked the music generated under these three settings. The questionnaire took an average of 29 minutes, and the average ranking results are shown in Table~\ref{tab:abla_ana}-(b). The results indicate that both the \textit{music-level} and \textit{bar-level} emotion control settings outperformed the \textit{no control} setting, indicating the importance of emotion control in generating background music for videos. Notably, the fine-grained emotion control at the bar level achieved the best results across all five metrics, especially for the \textit{Emotion-Matching} metric, demonstrating the effectiveness of conducting fine-grained emotion control at the bar level. This strategy can capture subtle emotional changes on the fly and generate music results that are more emotionally aligned with the input video.

\subsubsection{The Controllability of Text and Image Prompts} To quantitatively showcase the controllability of text and image prompts, we evaluated the emotion classification accuracy on the self-constructed test sets described in Sec.~\ref{sec:exp_dataset}. The image-based emotion classification task achieved an accuracy of 85.2\%, while the text-based task reached 87.7\%. This high performance can be attributed to the integration of general emotional knowledge from large-scale models such as CLIP~\cite{radford2021learning} and SentenceTransformer~\cite{reimers-2019-sentence-bert}. As demonstrated in Sec.~\ref{sec:emo_subj}, given emotion tags, our \composer surpasses the current methods in emotion control capabilities. Thus, \projector accurately analyzes emotions from text and images, while \composer effectively generates emotion-specific music, demonstrating the strong controllability of \framework for text and image inputs.

\begin{table}[!t]
\caption{Ablation analysis on the weighting factors $\lambda_1$ and $\lambda_2$ of image emotion recognition task.}
\vspace{-5pt}
\label{tab:abla_image_emo_cls}
\begin{center}
\begin{tabular}{cccccccc}
\toprule
\textbf{$(\lambda_1,\lambda_2)$} & (0,1) & (1,0) & (1,1) & (1,2) & (1,3) & (2,1) & (3,1)\\
\midrule
\textbf{Accuracy (\%)} & 80.2 & 77.9 & 83.7 & \textbf{85.2} & 84.8 & 82.5 & 81.7 \\
\bottomrule
\end{tabular}
\vspace{-10pt}
\end{center}
\end{table}

\subsubsection{Ablation on the Weighting Factors of Image Emotion Recognition Task} The weighting factors $\lambda_{1}$ and $\lambda_{2}$ in Eqn.~\ref{equation:image} balance the contributions of two models (\textit{i.e.}, ResNet and CLIP). We explored various numerical configurations to evaluate their impact on the image emotion recognition task. As shown in Table~\ref{tab:abla_image_emo_cls}, we observed that the single model settings ($\lambda_{1}=0$ or $\lambda_{2}=0$) were inferior to the dual-model consensus settings. Moreover, giving higher weight to the CLIP model ($\lambda_{1} < \lambda_{2}$) led to better results. Therefore, we set $\lambda_{1}=1$ and $\lambda_{2}=2$ as the default settings because they yielded the best overall performance.

\subsubsection{The Effectiveness of the Proposed Music Representation}\label{sec:effect_rep} As described in Sec.~\ref{sec:composer}, we designed an enhanced symbolic music representation based on the CP~\cite{hsiao2021compound} representation, which includes three key family token improvements: Tag, Instrument, and Rhythm. To evaluate the efficacy of these new family tokens, we conducted a series of ablation studies on the XMIDI dataset. Starting from the baseline CP representation, we incrementally added Tag, Instrument, and Rhythm family tokens to investigate their impact on music generation. The participants in the subjective evaluation ranked the generated music based on three aspects: richness, correctness, and structuredness. The average time to complete the questionnaire was 83 minutes, and the results are summarized in Table~\ref{tab:abla_ana}-(c). We can conclude that each added family token significantly enhanced performance, demonstrating the effectiveness of the proposed symbolic music representation.

\subsection{Discussion}

\subsubsection{Data Distribution of the XMIDI Dataset} We designed a comparative experiment to investigate whether the imbalance issue in XMIDI dataset affects the quality of the generated music. The two common strategies for addressing imbalanced data are \textit{undersampling} common classes and \textit{oversampling} rare classes through duplication. We applied these two sampling strategies to balance the data categories and trained models with an equal number of iterations. We then conducted a subjective evaluation comparing the performance of these models with a baseline model trained on the original, imbalanced dataset. The participants ranked the generated music pieces from these three models, and the questionnaire took approximately 61 minutes to complete. As shown in Table~\ref{tab:abla_ana}-(d), we have the following observations: i) The model trained with the undersampling strategy performed worse than the one trained on the original XMIDI dataset, demonstrating that data diversity is more critical than balance for music generation; ii) The oversampling strategy was also inferior to the baseline, likely due to overfitting issue caused by simple data duplication; iii) The model trained on the original XMIDI dataset achieved the highest performance. We conjecture that the diverse data sources of our XMIDI reflects the long-tailed distribution of real-world music categories, better aligns with human auditory preferences. In future work, we will expand the number of rare categories or explore more effective augmentation strategies to further address the data imbalance challenge effectively.

\subsubsection{Limitations and Future Work} Our XMusic supports five common input modalities for controllable music generation: videos, images, texts, tags, and humming. Other modalities, such as human skeletons, gestures, and depth, are worth further exploration. Additionally, this paper focuses on a subset of symbolic music elements, while a broader range of elements, such as time signatures, music lengths, and keys, could be incorporated for more comprehensive control. Currently, XMusic analyzes only the global emotion expressed in the text prompt and does not explicitly consider specific music elements mentioned within the text. To address this, we plan to train a text classifier in the future to better extract the music elements contained in the text for more precise control. Moreover, we aim to further expand the XMIDI dataset, particularly for rare emotion and genre categories, to build a more balanced and larger-scale symbolic music dataset.

\section{Conclusion}
In this paper, we propose a multi-modal controllable symbolic music generation framework called \framework. This framework supports versatile prompts, such as videos, images, texts, tags, and humming. Music elements act as connectors between the prompt parsing and generation controlling processes, explicitly decoupling the control signal analysis task from the music generation pipeline. This design enjoys strong scalability, facilitating the integration of new modalities in a plug-and-play manner. Specifically, our \projector parses multi-modal prompts into symbolic music elements within the projection space, while \composer generates high-quality music aligned with the control conditions based on our enhanced symbolic music representation. Furthermore, we construct a large-scale symbolic music dataset called \dataset with precise emotion and genre annotations for training the music generation model. Compared to the current state-of-the-art methods, \framework achieves superior performance across all utilized objective and subjective evaluation metrics.

\bibliographystyle{IEEEtran}
% argument is your BibTeX string definitions and bibliography database(s)
% \bibliography{IEEEabrv,references.bib}

\begin{thebibliography}{10}
\providecommand{\url}[1]{#1}
\csname url@samestyle\endcsname
\providecommand{\newblock}{\relax}
\providecommand{\bibinfo}[2]{#2}
\providecommand{\BIBentrySTDinterwordspacing}{\spaceskip=0pt\relax}
\providecommand{\BIBentryALTinterwordstretchfactor}{4}
\providecommand{\BIBentryALTinterwordspacing}{\spaceskip=\fontdimen2\font plus
\BIBentryALTinterwordstretchfactor\fontdimen3\font minus
  \fontdimen4\font\relax}
\providecommand{\BIBforeignlanguage}[2]{{%
\expandafter\ifx\csname l@#1\endcsname\relax
\typeout{** WARNING: IEEEtran.bst: No hyphenation pattern has been}%
\typeout{** loaded for the language `#1'. Using the pattern for}%
\typeout{** the default language instead.}%
\else
\language=\csname l@#1\endcsname
\fi
#2}}
\providecommand{\BIBdecl}{\relax}
\BIBdecl

\bibitem{borsos2023audiolm}
Z.~Borsos, R.~Marinier, D.~Vincent, E.~Kharitonov, O.~Pietquin, M.~Sharifi,
  D.~Roblek, O.~Teboul, D.~Grangier, M.~Tagliasacchi \emph{et~al.}, ``Audiolm:
  a language modeling approach to audio generation,'' \emph{IEEE/ACM
  Transactions on Audio, Speech, and Language Processing}, 2023.

\bibitem{agostinelli2023musiclm}
A.~Agostinelli, T.~I. Denk, Z.~Borsos, J.~Engel, M.~Verzetti, A.~Caillon,
  Q.~Huang, A.~Jansen, A.~Roberts, M.~Tagliasacchi \emph{et~al.}, ``Musiclm:
  Generating music from text,'' \emph{arXiv preprint arXiv:2301.11325}, 2023.

\bibitem{Forsgren_Martiros_2022}
\BIBentryALTinterwordspacing
S.~Forsgren and H.~Martiros, ``{Riffusion - Stable diffusion for real-time
  music generation},'' 2022. [Online]. Available:
  \url{https://riffusion.com/about}
\BIBentrySTDinterwordspacing

\bibitem{copet2023simple}
J.~Copet, F.~Kreuk, I.~Gat, T.~Remez, D.~Kant, G.~Synnaeve, Y.~Adi, and
  A.~D{\'e}fossez, ``Simple and controllable music generation,'' \emph{arXiv
  preprint arXiv:2306.05284}, 2023.

\bibitem{huang2023noise2music}
Q.~Huang, D.~S. Park, T.~Wang, T.~I. Denk, A.~Ly, N.~Chen, Z.~Zhang, Z.~Zhang,
  J.~Yu, C.~Frank \emph{et~al.}, ``Noise2music: Text-conditioned music
  generation with diffusion models,'' \emph{arXiv preprint arXiv:2302.03917},
  2023.

\bibitem{vaswani2017attention}
A.~Vaswani, N.~Shazeer, N.~Parmar, J.~Uszkoreit, L.~Jones, A.~N. Gomez,
  {\L}.~Kaiser, and I.~Polosukhin, ``Attention is all you need,''
  \emph{Advances in neural information processing systems}, vol.~30, 2017.

\bibitem{dai2019transformer}
Z.~Dai, Z.~Yang, Y.~Yang, J.~G. Carbonell, Q.~Le, and R.~Salakhutdinov,
  ``Transformer-xl: Attentive language models beyond a fixed-length context,''
  in \emph{Proceedings of the 57th Annual Meeting of the Association for
  Computational Linguistics}, 2019, pp. 2978--2988.

\bibitem{huangmusic}
C.-Z.~A. Huang, A.~Vaswani, J.~Uszkoreit, I.~Simon, C.~Hawthorne, N.~Shazeer,
  A.~M. Dai, M.~D. Hoffman, M.~Dinculescu, and D.~Eck, ``Music transformer:
  Generating music with long-term structure,'' in \emph{International
  Conference on Learning Representations}.

\bibitem{huang2020pop}
Y.-S. Huang and Y.-H. Yang, ``Pop music transformer: Beat-based modeling and
  generation of expressive pop piano compositions,'' in \emph{Proceedings of
  the 28th ACM International Conference on Multimedia}, 2020, pp. 1180--1188.

\bibitem{hsiao2021compound}
W.-Y. Hsiao, J.-Y. Liu, Y.-C. Yeh, and Y.-H. Yang, ``Compound word transformer:
  Learning to compose full-song music over dynamic directed hypergraphs,'' in
  \emph{Proceedings of the AAAI Conference on Artificial Intelligence},
  vol.~35, no.~1, 2021, pp. 178--186.

\bibitem{EMOPIA}
H.-T. Hung, J.~Ching, S.~Doh, N.~Kim, J.~Nam, and Y.-H. Yang, ``{EMOPIA}: A
  multi-modal pop piano dataset for emotion recognition and emotion-based music
  generation,'' in \emph{Proc. Int. Society for Music Information Retrieval
  Conf.}, 2021.

\bibitem{bao2022generating}
C.~Bao and Q.~Sun, ``Generating music with emotions,'' \emph{IEEE Transactions
  on Multimedia}, 2022.

\bibitem{payne2019musenet}
C.~Payne, ``Musenet,'' \emph{OpenAI Blog}, vol.~3, 2019.

\bibitem{sarmento2023gtr}
P.~Sarmento, A.~Kumar, Y.-H. Chen, C.~Carr, Z.~Zukowski, and M.~Barthet,
  ``Gtr-ctrl: Instrument and genre conditioning for guitar-focused music
  generation with transformers,'' in \emph{Artificial Intelligence in Music,
  Sound, Art and Design: 12th International Conference, EvoMUSART 2023, Held as
  Part of EvoStar 2023, Brno, Czech Republic, April 12--14, 2023,
  Proceedings}.\hskip 1em plus 0.5em minus 0.4em\relax Springer, 2023, pp.
  260--275.

\bibitem{EmoMusicTV}
S.~Ji and X.~Yang, ``Emomusictv: Emotion-conditioned symbolic music generation
  with hierarchical transformer vae,'' \emph{IEEE Transactions on Multimedia},
  pp. 1--13, 2023.

\bibitem{Zou2021MelonsGM}
Y.~Zou, P.~Zou, Y.~Zhao, K.~Zhang, R.~Zhang, and X.~Wang, ``Melons: Generating
  melody with long-term structure using transformers and structure graph,''
  \emph{ICASSP 2022 - 2022 IEEE International Conference on Acoustics, Speech
  and Signal Processing (ICASSP)}, pp. 191--195, 2021.

\bibitem{gan2020foley}
C.~Gan, D.~Huang, P.~Chen, J.~B. Tenenbaum, and A.~Torralba, ``Foley music:
  Learning to generate music from videos,'' in \emph{Computer Vision--ECCV
  2020: 16th European Conference, Glasgow, UK, August 23--28, 2020,
  Proceedings, Part XI 16}.\hskip 1em plus 0.5em minus 0.4em\relax Springer,
  2020, pp. 758--775.

\bibitem{su2020audeo}
K.~Su, X.~Liu, and E.~Shlizerman, ``Audeo: Audio generation for a silent
  performance video,'' \emph{Advances in Neural Information Processing
  Systems}, vol.~33, pp. 3325--3337, 2020.

\bibitem{di2021video}
S.~Di, Z.~Jiang, S.~Liu, Z.~Wang, L.~Zhu, Z.~He, H.~Liu, and S.~Yan, ``Video
  background music generation with controllable music transformer,'' in
  \emph{Proceedings of the 29th ACM International Conference on Multimedia},
  2021, pp. 2037--2045.

\bibitem{zhuo2023video}
L.~Zhuo, Z.~Wang, B.~Wang, Y.~Liao, C.~Bao, S.~Peng, S.~Han, A.~Zhang, F.~Fang,
  and S.~Liu, ``Video background music generation: Dataset, method and
  evaluation,'' in \emph{Proceedings of the IEEE/CVF International Conference
  on Computer Vision}, 2023, pp. 15\,637--15\,647.

\bibitem{ouyang2022training}
L.~Ouyang, J.~Wu, X.~Jiang, D.~Almeida, C.~Wainwright, P.~Mishkin, C.~Zhang,
  S.~Agarwal, K.~Slama, A.~Ray \emph{et~al.}, ``Training language models to
  follow instructions with human feedback,'' \emph{Advances in Neural
  Information Processing Systems}, vol.~35, pp. 27\,730--27\,744, 2022.

\bibitem{OpenAI2023GPT4TR}
OpenAI, ``Gpt-4 technical report,'' \emph{ArXiv}, vol. abs/2303.08774, 2023.

\bibitem{thoppilan2022lamda}
R.~Thoppilan, D.~De~Freitas, J.~Hall, N.~Shazeer, A.~Kulshreshtha, H.-T. Cheng,
  A.~Jin, T.~Bos, L.~Baker, Y.~Du \emph{et~al.}, ``Lamda: Language models for
  dialog applications,'' \emph{arXiv preprint arXiv:2201.08239}, 2022.

\bibitem{karras2019style}
T.~Karras, S.~Laine, and T.~Aila, ``A style-based generator architecture for
  generative adversarial networks,'' in \emph{Proceedings of the IEEE/CVF
  conference on computer vision and pattern recognition}, 2019, pp. 4401--4410.

\bibitem{ramesh2022hierarchical}
A.~Ramesh, P.~Dhariwal, A.~Nichol, C.~Chu, and M.~Chen, ``Hierarchical
  text-conditional image generation with clip latents,'' \emph{arXiv preprint
  arXiv:2204.06125}, 2022.

\bibitem{rombach2022high}
R.~Rombach, A.~Blattmann, D.~Lorenz, P.~Esser, and B.~Ommer, ``High-resolution
  image synthesis with latent diffusion models,'' in \emph{Proceedings of the
  IEEE/CVF Conference on Computer Vision and Pattern Recognition}, 2022, pp.
  10\,684--10\,695.

\bibitem{schneider2023mo}
F.~Schneider, Z.~Jin, and B.~Sch{\"o}lkopf, ``Mo\^{u}sai: Text-to-music
  generation with long-context latent diffusion,'' \emph{arXiv preprint
  arXiv:2301.11757}, 2023.

\bibitem{Yang2017MidiNetAC}
L.-C. Yang, S.-Y. Chou, and Y.-H. Yang, ``Midinet: A convolutional generative
  adversarial network for symbolic-domain music generation,'' in
  \emph{International Society for Music Information Retrieval Conference},
  2017.

\bibitem{dong2018musegan}
H.-W. Dong, W.-Y. Hsiao, L.-C. Yang, and Y.-H. Yang, ``Musegan: Multi-track
  sequential generative adversarial networks for symbolic music generation and
  accompaniment,'' in \emph{Proceedings of the AAAI Conference on Artificial
  Intelligence}, vol.~32, no.~1, 2018.

\bibitem{choi2020encoding}
K.~Choi, C.~Hawthorne, I.~Simon, M.~Dinculescu, and J.~Engel, ``Encoding
  musical style with transformer autoencoders,'' in \emph{International
  Conference on Machine Learning}.\hskip 1em plus 0.5em minus 0.4em\relax PMLR,
  2020, pp. 1899--1908.

\bibitem{jiang2020transformer}
J.~Jiang, G.~G. Xia, D.~B. Carlton, C.~N. Anderson, and R.~H. Miyakawa,
  ``Transformer vae: A hierarchical model for structure-aware and interpretable
  music representation learning,'' in \emph{ICASSP 2020-2020 IEEE International
  Conference on Acoustics, Speech and Signal Processing (ICASSP)}.\hskip 1em
  plus 0.5em minus 0.4em\relax IEEE, 2020, pp. 516--520.

\bibitem{brunner2018symbolic}
G.~Brunner, Y.~Wang, R.~Wattenhofer, and S.~Zhao, ``Symbolic music genre
  transfer with cyclegan,'' in \emph{2018 ieee 30th international conference on
  tools with artificial intelligence (ictai)}.\hskip 1em plus 0.5em minus
  0.4em\relax IEEE, 2018, pp. 786--793.

\bibitem{Dong2018ConvolutionalGA}
H.-W. Dong and Y.-H. Yang, ``Convolutional generative adversarial networks with
  binary neurons for polyphonic music generation,'' in \emph{International
  Society for Music Information Retrieval Conference}, 2018.

\bibitem{sdmuse}
C.~Zhang, Y.~Ren, K.~Zhang, and S.~Yan, ``Sdmuse: Stochastic differential music
  editing and generation via hybrid representation,'' \emph{IEEE Transactions
  on Multimedia}, pp. 1--9, 2023.

\bibitem{Dai2021ControllableDM}
S.~Dai, Z.~Jin, C.~Gomes, and R.~B. Dannenberg, ``Controllable deep melody
  generation via hierarchical music structure representation,'' in
  \emph{International Society for Music Information Retrieval Conference},
  2021.

\bibitem{Medeot2018StructureNetIS}
G.~Medeot, S.~Cherla, K.~Kosta, M.~McVicar, S.~M. Abdallah, M.~Selvi,
  E.~Newton-Rex, and K.~Webster, ``Structurenet: Inducing structure in
  generated melodies,'' in \emph{International Society for Music Information
  Retrieval Conference}, 2018.

\bibitem{jhamtani2019modeling}
H.~Jhamtani and T.~Berg-Kirkpatrick, ``Modeling self-repetition in music
  generation using generative adversarial networks,'' in \emph{Machine Learning
  for Music Discovery Workshop, ICML}, 2019.

\bibitem{herremans2017morpheus}
D.~Herremans and E.~Chew, ``Morpheus: generating structured music with
  constrained patterns and tension,'' \emph{IEEE Transactions on Affective
  Computing}, vol.~10, no.~4, pp. 510--523, 2017.

\bibitem{neves2022generating}
P.~Neves, J.~Fornari, and J.~Florindo, ``Generating music with sentiment using
  transformer-gans,'' \emph{arXiv preprint arXiv:2212.11134}, 2022.

\bibitem{MGM2023}
Z.~Hu, X.~Ma, Y.~Liu, G.~Chen, Y.~Liu, and R.~B. Dannenberg, ``The beauty of
  repetition: an algorithmic composition model with motif-level repetition
  generator and outline-to-music generator in symbolic music generation,''
  \emph{IEEE Transactions on Multimedia}, pp. 1--14, 2023.

\bibitem{shih2022theme}
Y.-J. Shih, S.-L. Wu, F.~Zalkow, M.~Muller, and Y.-H. Yang, ``Theme
  transformer: Symbolic music generation with theme-conditioned transformer,''
  \emph{IEEE Transactions on Multimedia}, 2022.

\bibitem{uptrans}
Z.~Hu, Y.~Liu, G.~Chen, and Y.~Liu, ``Can machines generate personalized music?
  a hybrid favorite-aware method for user preference music transfer,''
  \emph{IEEE Transactions on Multimedia}, vol.~25, pp. 2296--2308, 2023.

\bibitem{yu2021conditional}
Y.~Yu, A.~Srivastava, and S.~Canales, ``Conditional lstm-gan for melody
  generation from lyrics,'' \emph{ACM Transactions on Multimedia Computing,
  Communications, and Applications (TOMM)}, vol.~17, no.~1, pp. 1--20, 2021.

\bibitem{zhang2024controllable}
Z.~Zhang, Y.~Yu, and A.~Takasu, ``Controllable syllable-level lyrics generation
  from melody with prior attention,'' \emph{IEEE Transactions on Multimedia},
  2024.

\bibitem{duan2023melody}
W.~Duan, Y.~Yu, X.~Zhang, S.~Tang, W.~Li, and K.~Oyama, ``Melody generation
  from lyrics with local interpretability,'' \emph{ACM Transactions on
  Multimedia Computing, Communications and Applications}, vol.~19, no.~3, pp.
  1--21, 2023.

\bibitem{NEURIPS2021_f4e369c0}
K.~Su, X.~Liu, and E.~Shlizerman, ``How does it sound?'' in \emph{Advances in
  Neural Information Processing Systems}, vol.~34, 2021, pp. 29\,258--29\,273.

\bibitem{zhu2022quantized}
Y.~Zhu, K.~Olszewski, Y.~Wu, P.~Achlioptas, M.~Chai, Y.~Yan, and S.~Tulyakov,
  ``Quantized gan for complex music generation from dance videos,'' in
  \emph{Computer Vision--ECCV 2022: 17th European Conference, Tel Aviv, Israel,
  October 23--27, 2022, Proceedings, Part XXXVII}.\hskip 1em plus 0.5em minus
  0.4em\relax Springer, 2022, pp. 182--199.

\bibitem{he2016deep}
K.~He, X.~Zhang, S.~Ren, and J.~Sun, ``Deep residual learning for image
  recognition,'' in \emph{Proceedings of the IEEE conference on computer vision
  and pattern recognition}, 2016, pp. 770--778.

\bibitem{radford2021learning}
A.~Radford, J.~W. Kim, C.~Hallacy, A.~Ramesh, G.~Goh, S.~Agarwal, G.~Sastry,
  A.~Askell, P.~Mishkin, J.~Clark \emph{et~al.}, ``Learning transferable visual
  models from natural language supervision,'' in \emph{International conference
  on machine learning}.\hskip 1em plus 0.5em minus 0.4em\relax PMLR, 2021, pp.
  8748--8763.

\bibitem{reimers-2019-sentence-bert}
N.~Reimers and I.~Gurevych, ``Sentence-bert: Sentence embeddings using siamese
  bert-networks,'' in \emph{Proceedings of the 2019 Conference on Empirical
  Methods in Natural Language Processing}.\hskip 1em plus 0.5em minus
  0.4em\relax Association for Computational Linguistics, 11 2019.

\bibitem{pyscenedetect}
\BIBentryALTinterwordspacing
``Pyscenedetect documentation.'' [Online]. Available:
  \url{https://scenedetect.com/en/latest/}
\BIBentrySTDinterwordspacing

\bibitem{zhang2019pan}
C.~Zhang, Y.~Zou, G.~Chen, and L.~Gan, ``Pan: Persistent appearance network
  with an efficient motion cue for fast action recognition,'' in
  \emph{Proceedings of the 27th ACM International Conference on Multimedia},
  2019, pp. 500--509.

\bibitem{davis2018visual}
A.~Davis and M.~Agrawala, ``Visual rhythm and beat,'' \emph{ACM Transactions on
  Graphics (TOG)}, vol.~37, no.~4, pp. 1--11, 2018.

\bibitem{hsu2021vocano}
J.-Y. Hsu and L.~Su, ``Vocano: A note transcription framework for singing voice
  in polyphonic music.'' in \emph{ISMIR}, 2021, pp. 293--300.

\bibitem{panda2013multi}
R.~Panda, R.~Malheiro, B.~Rocha, A.~Oliveira, and R.~P. Paiva, ``Multi-modal
  music emotion recognition: A new dataset, methodology and comparative
  analysis,'' in \emph{International symposium on computer music
  multidisciplinary research}, 2013.

\bibitem{ferreira-2019-learning}
L.~Ferreira and J.~Whitehead, ``Learning to generate music with sentiment,'' in
  \emph{Proceedings of the 20th International Society for Music Information
  Retrieval Conference, {ISMIR} 2019, Delft, The Netherlands, November 4-8,
  2019}, 2019, pp. 384--390.

\bibitem{panda2018contemplating}
R.~Panda, J.~Zhang, H.~Li, J.-Y. Lee, X.~Lu, and A.~K. Roy-Chowdhury,
  ``Contemplating visual emotions: Understanding and overcoming dataset bias,''
  in \emph{European Conference on Computer Vision}, 2018.

\bibitem{zhou2017places}
B.~Zhou, A.~Lapedriza, A.~Khosla, A.~Oliva, and A.~Torralba, ``Places: A 10
  million image database for scene recognition,'' \emph{IEEE Transactions on
  Pattern Analysis and Machine Intelligence}, 2017.

\bibitem{Wu2020TheJT}
S.-L. Wu and Y.-H. Yang, ``The jazz transformer on the front line: Exploring
  the shortcomings of ai-composed music through quantitative measures,'' in
  \emph{International Society for Music Information Retrieval Conference},
  2020.

\bibitem{dong2018pypianoroll}
H.-W. Dong, W.-Y. Hsiao, and Y.-H. Yang, ``Pypianoroll: Open source python
  package for handling multitrack pianoroll,'' \emph{Proc. ISMIR. Late-breaking
  paper}, 2018.

\bibitem{dong2020muspy}
H.-W. Dong, K.~Chen, J.~McAuley, and T.~Berg-Kirkpatrick, ``Muspy: A toolkit
  for symbolic music generation,'' \emph{arXiv preprint arXiv:2008.01951},
  2020.

\bibitem{wu2022exploring}
S.~Wu and M.~Sun, ``Exploring the efficacy of pre-trained checkpoints in
  text-to-music generation task,'' \emph{arXiv preprint arXiv:2211.11216},
  2022.

\bibitem{tan2020automated}
X.~Tan, M.~Antony, and H.~Kong, ``Automated music generation for visual art
  through emotion.'' in \emph{ICCC}, 2020, pp. 247--250.

\bibitem{bubeck2023sparks}
S.~Bubeck, V.~Chandrasekaran, R.~Eldan, J.~Gehrke, E.~Horvitz, E.~Kamar,
  P.~Lee, Y.~T. Lee, Y.~Li, S.~Lundberg \emph{et~al.}, ``Sparks of artificial
  general intelligence: Early experiments with gpt-4,'' \emph{arXiv preprint
  arXiv:2303.12712}, 2023.

\bibitem{wu2008study}
X.~Wu, ``A study on image-based music generation,'' 2008.

\bibitem{madhok2018sentimozart}
R.~Madhok, S.~Goel, and S.~Garg, ``Sentimozart: Music generation based on
  emotions.'' in \emph{ICAART (2)}, 2018, pp. 501--506.

\bibitem{qiu2023humming2music}
Y.~Qiu, J.~Zhang, H.~Ren, Y.~Shan, and J.~Zhou, ``Humming2music: being a
  composer as long as you can humming,'' in \emph{Proceedings of the
  Thirty-Second International Joint Conference on Artificial Intelligence},
  2023, pp. 7163--7166.

\end{thebibliography}
% Generated by IEEEtran.bst, version: 1.14 (2015/08/26)

\vfill

\end{document}